\documentclass[sigconf]{acmart}

\usepackage{enumitem}
\usepackage{listings}
\usepackage{multirow}
\usepackage{twemojis}
\usepackage{makecell}
\usepackage{setspace}

\AtBeginDocument{%
  }

\newcommand{\pipelinename}{\textsc{Personalized Reflect-Respond}}
\newcommand{\sysname}{\textsc{TeachTune}}
\newcommand\variable[1]{\textcolor{blue}{#1}}
\newcommand\generated[1]{\textcolor{orange}{#1}}
\newcommand\change[1]{\textcolor{red}{#1}}

\copyrightyear{2025}
\acmYear{2025}
\setcopyright{cc}
\setcctype{by-nc-sa}
\acmConference[CHI '25]{CHI Conference on Human Factors in Computing Systems}{April 26-May 1, 2025}{Yokohama, Japan}
\acmBooktitle{CHI Conference on Human Factors in Computing Systems (CHI '25), April 26-May 1, 2025, Yokohama, Japan}\acmDOI{10.1145/3706598.3714054}
\acmISBN{979-8-4007-1394-1/25/04}

\begin{document}

\title{\sysname{}: Reviewing Pedagogical Agents Against Diverse Student Profiles with Simulated Students}

\author{Hyoungwook Jin}
\email{jinhw@kaist.ac.kr}
\orcid{0000-0003-0253-560X}
\affiliation{%
  \institution{School of Computing, KAIST}
  \city{Daejeon}
  \country{Republic of Korea}
}

\author{Minju Yoo}
\email{minjuu613@ewhain.net}
\orcid{0009-0000-2964-0464}
\affiliation{%
  \institution{Ewha Womans University}
  \city{Seoul}
  \country{Republic of Korea}
}

\author{Jeongeon Park}
\email{jep034@ucsd.edu}
\orcid{0000-0002-8353-0431}
\affiliation{%
  \institution{University of California San Diego}
  \city{La Jolla}
  \country{United States}
}

\author{Yokyung Lee}
\email{ykleeee@kaist.ac.kr}
\orcid{0009-0000-8732-9316}
\affiliation{%
  \institution{School of Computing, KAIST}
  \city{Daejeon}
  \country{Republic of Korea}
}

\author{Xu Wang}
\email{xwanghci@umich.edu}
\orcid{0000-0001-5551-0815}
\affiliation{%
  \institution{Computer Science and Engineering, University of Michigan}
  \city{Ann Arbor}
  \country{United States}
}

\author{Juho Kim}
\email{juhokim@kaist.ac.kr}
\orcid{0000-0001-6348-4127}
\affiliation{%
  \institution{School of Computing, KAIST}
  \city{Daejeon}
  \country{Republic of Korea}
}


\begin{abstract}
Large language models (LLMs) can empower teachers to build pedagogical conversational agents (PCAs) customized for their students. As students have different prior knowledge and motivation levels, teachers must review the adaptivity of their PCAs to diverse students. Existing chatbot reviewing methods (e.g., direct chat and benchmarks) are either manually intensive for multiple iterations or limited to testing only single-turn interactions. We present \sysname{}, where teachers can create simulated students and review PCAs by observing automated chats between PCAs and simulated students. Our technical pipeline instructs an LLM-based student to simulate prescribed knowledge levels and traits, helping teachers explore diverse conversation patterns. Our pipeline could produce simulated students whose behaviors correlate highly to their input knowledge and motivation levels within 5\% and 10\% accuracy gaps. Thirty science teachers designed PCAs in a between-subjects study, and using \sysname{} resulted in a lower task load and higher student profile coverage over a baseline.
\end{abstract}



\begin{teaserfigure}
  \includegraphics[width=\textwidth]{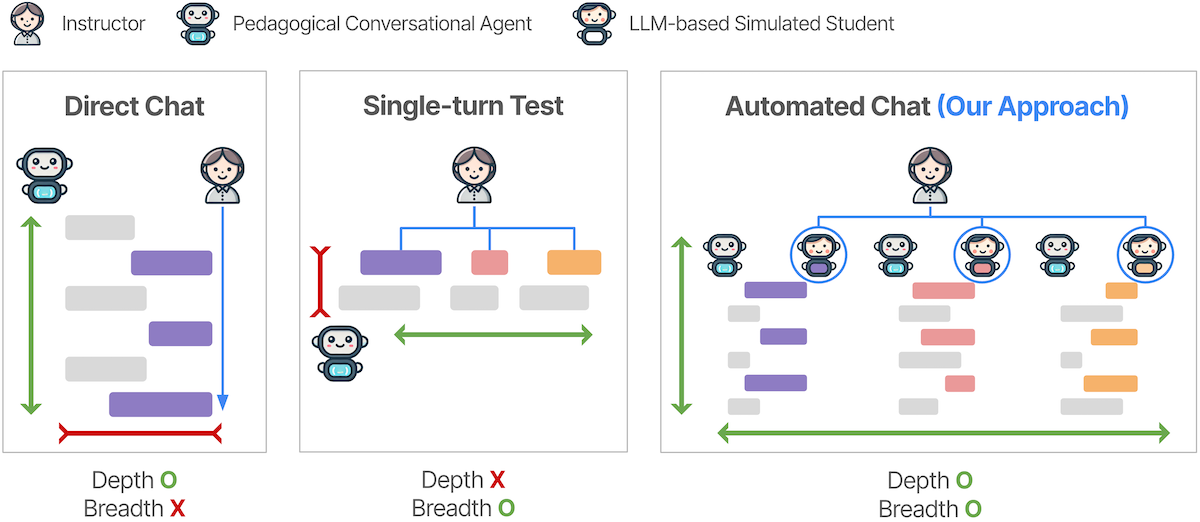}
  \caption{\sysname{} is an evaluation tool that helps teachers review the interaction quality of pedagogical agents by utilizing simulated students. Direct chat supports an in-depth assessment but in a narrow scope. Single-turn tests with benchmark datasets support breadth exploration of pedagogical agents' adaptivity but lack depth in assessing conversations in multi-turn. \sysname{} takes the best of both worlds by leveraging automated chat between the pedagogical agent and user-defined simulated students to help teachers review the adaptivity of pedagogical agents in sufficient depth and breadth.}
  \Description{A figure comparing three educational methods: Direct Chat, Single-turn Test, and Automated Chat. Direct Chat offers breadth, benchmark dataset offers depth, and Automated Chat achieves both depth and breadth by using simulated agents.}
  \label{fig:teaser}
\end{teaserfigure}


\maketitle

\section{Introduction}

\begin{quote}
    ``\textit{A key challenge in developing and deploying Machine Learning (ML) systems is understanding their performance across a wide range of inputs.}''~\cite{wexler2019if}
\end{quote}

Large Language Models (LLMs) have empowered teachers to build Pedagogical Conversational Agents (PCAs)~\cite{weber2021pedagogical} with little programming expertise. 
\textbf{PCAs} refer to conversational agents that act as instructors~\cite{graesser2004autotutor}, peers~\cite{matsuda2012studying}, and motivators~\cite{alaimi2020pedagogical} with whom learners can communicate through natural language, used in diverse subjects, grades, and pedagogies.
Teacher-designed PCAs can better adapt to downstream class environments (i.e., students and curriculum) and allow teachers to experiment with diverse class activities that were previously prohibitive due to limited human resources. 
While conventional chatbots require authoring hard-coded conversational flows and responses~\cite{cranshaw2017calendar, choi2021protochat}, LLM-based agents need only a description of how the agents should behave in natural language, known as prompting~\cite{dam2024complete}. Prior research has proposed prompting techniques~\cite{wu2022ai, brown2020language, yao2023tree}, user interfaces~\cite{arawjo2024chainforge, martin2024chedbot, fiannaca2023programming}, and frameworks~\cite{kim2023cells, bai2022constitutional} that make domain-specific and personalized agents even more accessible to build for end-users. With the lowered barrier and cost of making conversational agents, researchers have actively experimented with LLM-based PCAs under diverse pedagogical settings, such as 1-on-1 tutoring~\cite{jurenka2024towards, zamfirescu2023conversational, han2023recipe}, peer learning~\cite{jin2024teach, schmucker2023}, and collaborative learning~\cite{liu2024peergpt, wei2024improving, nguyen2024simulating}.

To disseminate these experimental PCAs to actual classes at scale, reviewing agents' content and interaction qualities is necessary before deployment. Many countries and schools are concerned about the potential harms of LLMs and hesitant about their use in classrooms, especially K-12, despite the benefits~\cite{johnson2023chatgpt, li2024chatgpt}. LLM-based PCAs need robust validation against hallucination~\cite{shoufan2023exploring, park2024promise}, social biases~\cite{warr2024implicit, wambsganss2023unraveling}, and overreliance~\cite{dwivedi2023opinion, mosaiyebzadeh2023exploring}. Moreover, since students vary in their levels of knowledge and learning attitudes in a class~\cite{richardson2012psychological}, teachers must review how well their PCAs can cover diverse students in advance to help each student improve attitudes and learn better~\cite{bloom19842, tian2021let, pereira2016leveraging, pereira2021struggling}. For instance, teachers should check whether PCAs help not only poorly performing students fill knowledge gaps but also well-performing students build further knowledge through discussions. 
Regarding students' personalities, teachers should check if PCAs ask questions to prompt inactive students and compliment active students to keep them motivated.
These attempts contribute to improving fairness in learning, closing the growth gap between students instead of widening it~\cite{memarian2023fairness}.

However, existing methods for reviewing the PCAs' coverage of various student profiles offer limited breadth and depth (Fig.~\ref{fig:teaser}). The current landscape of chatbot evaluation takes two approaches at large. First, teachers can directly chat with their PCAs and roleplay themselves as students~\cite{petridis2024constitutionmaker, hedderich2024piece, bubeck2023sparks}. Although interactive chats allow teachers to review the behaviors of PCAs over multi-turn conversations in depth, it is time-consuming for teachers to manually write messages and re-run conversations after revising PCA designs, restraining the breadth of reviewing different students. Second, teachers can simultaneously author many input messages as test cases (e.g., benchmark datasets) and assess the PCAs' responses~\cite{ribeiro2022adaptive, wu2023scattershot, zamfirescu2023johnny, cabrera2023zeno, kim2024evallm}. Single-turn test cases are scalable and reproducible, but teachers can examine only limited responses that do not capture multi-turn interactions (e.g., splitting explanations~\cite{lee2023dapie}, asking follow-up questions~\cite{shahriar2021can}), restricting the depth of each review. Teachers may also need to create test cases manually if their PCAs target new curriculums and class activities.

To support efficient PCA reviewing with breadth and depth, we propose a novel review method in which teachers utilize auto-generated conversations between a PCA and simulated students. Recent research has found that LLMs can simulate human behaviors of diverse personalities~\cite{li2024evolving, park2023generative} and knowledge levels~\cite{lu2024generative, jin2024teach}. We extend this idea to PCA review by simulating conversations between PCAs and students with LLM. We envision simulated conversations making PCA evaluation as reproducible and efficient as the test case approach while maintaining the benefit of reviewing multi-turn interactions like direct chat. Teachers can review the adaptivity of PCAs by configuring diverse simulated students as a unit of testing and examine the quality of interaction in depth by observing auto-generated conversations among them. We implemented this idea into \sysname{}, a tool that allows teachers to design PCAs and review their robustness against diverse students and multi-turn scenarios through automated chats with simulated students. Teachers can configure simulated students by adding or removing knowledge components and adjusting the intensity of student traits, such as self-efficacy and motivation. Our LLM-prompting pipeline, \pipelinename{}, takes configurations on knowledge and trait intensity levels (5-point scale) as inputs and generates a comprehensive overview to instruct simulated students to generate believable responses.

To evaluate the performance of \pipelinename{} in simulating targeted student behaviors, we asked ten teachers to interact with nine simulated students of varying knowledge and trait levels in a blinded condition and to predict the simulated students' configuration levels for knowledge and traits. We measured the difference between teacher-predicted and initially configured levels. Our pipeline showed a 5\% median error for knowledge components and a 10\% median error for student traits, implying that our simulated students' behaviors closely align with the expectations of teachers who configure them. 
We also conducted a between-subjects study with 30 teachers to evaluate how \sysname{} can help teachers efficiently review the interaction quality of PCAs in depth and breadth.
Study participants created and reviewed PCAs for middle school science classes using \sysname{} or a baseline system where PCA review was possible through only direct chats and single-turn test cases. We found that automated chats significantly help teachers explore a broader range of students within traits (large effect size, $\eta^2$=0.304) at a lower task load ($\eta^2$=0.395). 

This paper makes the following contributions:
\begin{itemize}
    \item \pipelinename{}, an LLM prompting pipeline that generates an overview of a target student's knowledge, motivation, and psychosocial context and follows the overview to simulate a believable student behavior.

    \item \sysname{}, an interface for teachers to efficiently review the coverage of PCAs against diverse knowledge levels and student traits.
    
    \item Empirical findings showing that \sysname{} can help teachers design PCAs at a lower task load and review more student profiles, compared to direct chats and test cases only.
\end{itemize}

\section{Related Work}
Our work aims to support the design and reviewing process of PCAs in diverse learning contexts. We outline the emergent challenges in designing conversational agents and how LLM-based simulation can tackle the problem.

\subsection{Conversational Agent Design Process}
Designing chatbots involves dedicated chatbot designers prototyping and then iteratively revising their designs through testing. Understanding and responding to a diverse range of potential user intents and needs is crucial to the chatbot's success. 
Popular methods include the Wizard-of-Oz approach to collect quality conversation data~\cite{klemmer2000suede} and co-design workshops to receive direct feedback from multiple stakeholders~\cite{chen2020creating, durall2023designing}.
Involving humans to simulate conversations or collecting feedback can help chatbot designers understand human-chatbot collaborative workflow~\cite{cranshaw2017calendar}, explore diverse needs of users~\cite{potts2021chatbots, candello2022unveiling}, or iterate their chatbot to handle edge cases~\cite{klemmer2000suede, choi2021protochat}. 
Typical chatbot reviewing methods include conducting a usability study with a defined set of chatbots' social characteristics~\cite{chaves2021should}, directly chatting 1-on-1 with the designed chatbot~\cite{petridis2024constitutionmaker}, and testing with domain experts~\cite{hedderich2024piece}. Such methods can yield quality evaluation but are costly as they need to be executed manually by humans. For more large-scale testing, designers can use existing test cases~\cite{ribeiro2022adaptive, bubeck2023sparks} or construct new test sets with LLMs~\cite{wu2023scattershot}. However, such evaluations happen in big chunks of single-turn conversations, which limits the depth of conversation dynamics throughout multiple turns. To complement the limitations, researchers have recently proposed leveraging LLMs as simulated users~\cite{de2023leveraging}, role-players~\cite{fang2024llm}, and agent authoring assistant~\cite{calo2024towards}. \sysname{} explores a similar thread of work in the context of education by utilizing simulated students to aid teachers' breadth- and depth-wise reviewing of PCAs.

\subsection{Simulating Human Behavior with LLMs}
Recent advancements in LLM have led researchers to explore the capabilities of LLMs in simulating humans and their environments, such as simulating psychology experiments~\cite{coda2024cogbench}, individuals' beliefs and preferences~\cite{namikoshi2024using, jiang2023personallm, shao2023character, chuang2023simulating, choi2024Proxona}, and social interactions~\cite{park2023generative, vezhnevets2023generative, li2024econagent, shaikh2024rehearsal}. 
In education, existing works have simulated student behaviors for testing learning contents~\cite{lu2024generative, he2024evaluating, nguyen2024simulating}, predicting cognitive states of students~\cite{xu2024eduagent, liu2024personality}, facilitating interactive pedagogy~\cite{jin2024teach}, and assisting teaching abilities of instructors~\cite{markel2023gpteach, zhang2024simulating, nguyen2024large, radmehr2024towards}. In deciding which specific attribute to simulate, existing simulation work has utilized either knowledge states~\cite{lu2024generative, jin2024teach, huang2024generating, nguyen2024simulating} or cognitive traits, such as personalities and mindset~\cite{markel2023gpteach, li2024evolving, wang2024patient}. 
However, simulating both knowledge states and personalities is necessary for authentic learning behaviors because cognitive traits, in addition to prior knowledge, are a strong indicator for predicting success in learning~\cite{besterfield1997characteristics, astin1992undergraduate, richardson2012psychological, chen2004cognitive, chrysafiadi2013student}.
Liu et al. explored utilizing cognitive and noncognitive aspects, such as the student's language proficiency and the Big Five personality, to simulate students at binary levels (e.g., low vs. high openness) for testing intelligent tutoring systems~\cite{liu2024personality}. Our work develops this idea further by presenting an LLM-powered pipeline that can configure and simulate both learners' knowledge and traits at a finer granularity (i.e., a five-point scale). Finer-grained control of student simulation will help teachers review PCAs against detailed student types, making their classes more inclusive.
\section{Formative Interview and Design Goals}
We conducted semi-structured interviews with five school teachers and observed how teachers review PCAs to investigate \textbf{RQ1}. More specifically, we aimed to gain a comprehensive understanding of what types of students teachers want PCAs to cover, what student traits (e.g., motivation level, stress) characterize those students, how teachers create student personas using those traits, and what challenges teachers have with existing PCA review methods (i.e., direct chat and test cases).

\begin{enumerate}
    \item[\textbf{RQ1:}] What are teachers' needs in reviewing PCAs and challenges in using direct chats and test cases?
\end{enumerate}

\subsection{Interviewees}
We recruited middle school science teachers through online teacher communities in Korea. We required teachers to possess either an education-related degree or at least one year of teaching experience. The teachers had diverse backgrounds (Table~\ref{table:formative_demographic}). The interview took place through Zoom for 1.5 hours, and interviewees were compensated KRW 50,000 (USD 38).

\begin{table*}[ht]
\begin{tabular}{cccccc}
\toprule
\multicolumn{1}{l}{\multirow{2}{*}{\textbf{Id}}} &
  \multicolumn{1}{l}{\multirow{2}{*}{\textbf{Period of teaching}}} &
  \multicolumn{1}{l}{\multirow{2}{*}{\textbf{Size of class}}} &
  \multicolumn{3}{c}{\textbf{Familiarity}} \\ \cline{4-6} 
\multicolumn{1}{l}{} &
  \multicolumn{1}{l}{} &
  \multicolumn{1}{l}{} &
  \textbf{Chatbots} &
  \textbf{Chatbot design process} &
  \textbf{ChatGPT} \\ \hline
I1 & 3 years  & 20 students  & Unfamiliar    & Very unfamiliar & Familiar      \\
I2 & 6 years  & 20 students  & Very familiar & Very familiar   & Very familiar \\
I3 & 16 years & 21 students  & Unfamiliar    & Very unfamiliar & Familiar      \\
I4 & 2 years  & 200 students & Very familiar & Familiar        & Very familiar \\
I5 & 1 year   & 90 students  & Familiar      & Familiar        & Unfamiliar    \\
\bottomrule
\end{tabular}%
\caption{Demographic information of the interview participants. We recruited five participants with varying levels of teaching experience, chatbot familiarity, chatbot design process, and ChatGPT familiarity. ``Very familiar'' indicates the frequent usage, ``Familiar'' for a little usage, ``Unfamiliar'' for passing knowledge, and ``Very unfamiliar'' for no knowledge.}
\Description{A table with four columns and six rows. On the header row are identifiers, period of teaching, class size, and familiarity (Chatbots, Chatbot design process, ChatGPT). The following rows describe what each participant's experiences were like for the headers.}
\label{table:formative_demographic}
\end{table*}

\begin{figure*}[ht]
\includegraphics[width=\textwidth]{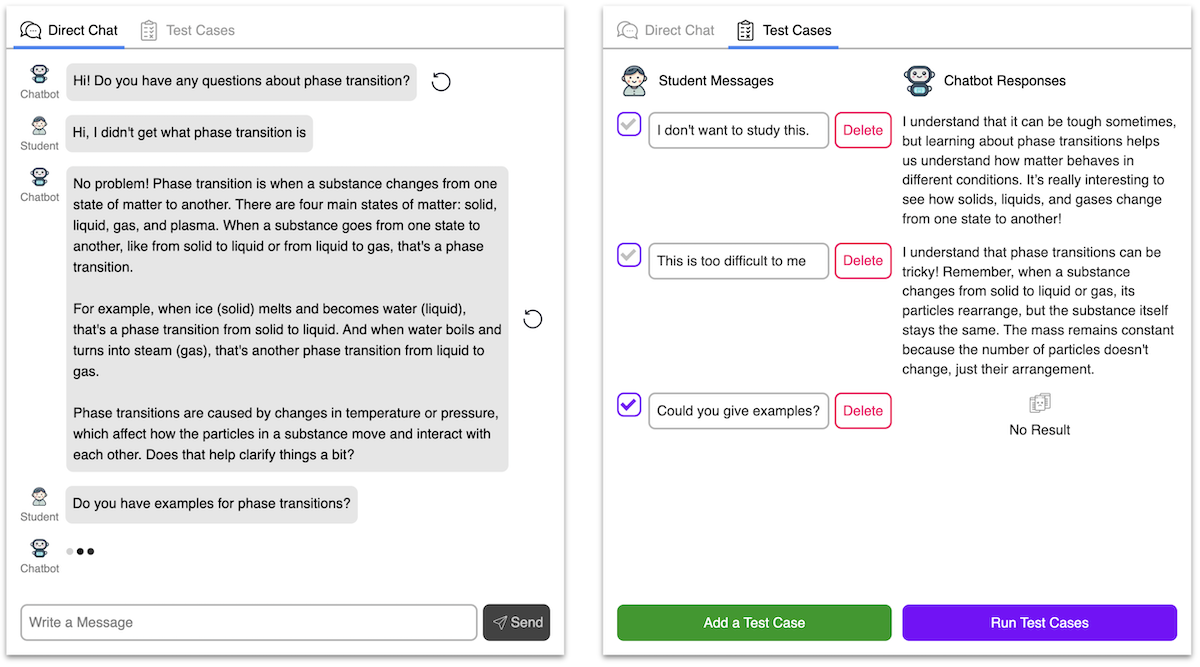}
    \caption{The interface used for the formative interview. On the left is the Direct Chat tab, where interviewees could converse with the chatbot as the student's role. Interviewees could roll back to previous messages by clicking the rewind button next to the chatbot's message. On the right is the Test Cases tab, where interviewees can add a set of student utterances and see chat responses.}
    \label{fig:formative_system}
    \Description{A screenshot showing the interface used for the formative interview. In the Direct Chat tab, the chatbot starts the conversation by saying, "Hi! Do you have any questions about phase transition?" The interviewee responded, "Hi, I didn't get what phase transition is," and the chatbot explained three states with real-life examples. In the Test Cases tab, the interviewee tested "I don't want to study this," "This is too difficult for me," and "Could you give examples."}
\end{figure*}

\subsection{Procedure}
We began the interview by presenting the research background, ChatGPT, and its various use cases (e.g., searching, brainstorming, and role-playing). We requested permission to record their voice and screen throughout the session and asked semi-structured interview questions during and after sessions.

Interviewees first identified the most critical student traits that PCAs should cover when supporting diverse students in K-12. To do so, we gave interviewees a list of 42 traits organized under five categories---personality traits, motivation factors, self-regulatory learning strategies, student approaches to learning, and psychosocial contextual influence~\cite{richardson2012psychological}. Interviewees ranked the categories by importance of reviewing and chose the top three traits from each category.

Interviewees then assumed a situation where they created PCAs for their science class to help students review the phase transition between solid, liquid, and gas.
Interviewees reviewed the interaction quality and adaptivity of a given tutor-role PCA by chatting with it directly and authoring test case messages, playing the role of students. Interviewees could revisit the list of 42 traits for their review. Interviewees used the interfaces in Fig.~\ref{fig:formative_system} for 10 minutes each and were asked to find as many limitations of the PCA as possible. The PCA was a GPT-3.5-based agent with the following system prompt: \textit{You are a middle school science teacher. You are having a conversation to help students understand what they learned in science class. Recently, students learned about phase transition. Help students if they have difficulty understanding phase transition.}

Subsequently, interviewees listed student profiles whose conversation with the PCA would help them review its quality and adaptivity. A student profile is distinguished from student traits as it is a combination of traits describing a student. Interviewees wrote student profiles in free form, using knowledge level and earlier 42 student traits to describe them (e.g., a student with average science grades but an introvert who prefers individual learning over cooperative learning). 

\begin{table*}[ht]
\begin{tabular}{lll}
\toprule
\textbf{Category}                   & \textbf{Student Trait}         & \textbf{Definition}                                                  \\ \hline
\multirow{2}{*}{Motivation factors} & Academic self-efficacy & Self-beliefs of academic capability                                  \\
                                    & Intrinsic motivation   & Inherent self-interest, and enjoyment of academic learning and tasks \\ \cline{2-3} 
\multirow{2}{*}{\begin{tabular}[c]{@{}l@{}}Psychosocial\\ contextual influence\end{tabular}} &
  Academic stress &
  Overwhelming negative emotionality resulting directly from academic stressors \\
                                    & Goal commitment        & Commitment to staying in school and obtaining a degree               \\ 
\bottomrule
\end{tabular}%
\caption{The top four student traits teachers found important for PCAs to cover.}
\label{table:student_traits}
\Description{Table displaying four traits by two categories with each definition: Academic self-efficacy, Intrinsic Motivation, Academic stress, and Goal commitment.}
\end{table*}

\subsection{Findings}\label{sec:interview_findings}
\subsubsection{Teachers deemed students' knowledge levels, motivation factors, and psychosocial contextual influences as important student traits to review.}
Interviewees thought that PCAs should support students with low motivation and knowledge, and hence, it is crucial to review how PCAs scaffold these students robustly. All five interviewees started their reviewing of the PCA with knowledge-related questions to assess the correctness and coverage of its knowledge. They then focused on how the PCA responds to a student with low motivation and interest (Table~\ref{table:student_traits}). Motivational factors (i.e., academic self-efficacy and intrinsic motivation) are important because students with low motivation often do not pay attention to class activities, and learning with a PCA would not work at all if the PCA cannot first encourage those students' participation (I1, I2, and I5). Interviewees also considered psychosocial factors (i.e., academic stress and goal commitment) important as they significantly affect the learning experience (I1). I3 remarked that she tried testing if the PCA could handle emotional questions because they take up most students' conversations. 

\subsubsection{Multi-turn conversations are crucial for review, but writing messages to converse with PCAs requires considerable effort and expertise.}
Follow-up questions and phased scaffolding are important pedagogical conversational patterns that appear over several message turns. Interviewees commented that it is critical to check how PCAs answer students' serial follow-up questions, use easier words across a conversation for struggling students, and remember conversational contexts because they affect learning and frequently happen in student-tutor conversations. Interviewees typically had 15 message turns for a comprehensive review of the PCA. Interviewees noted that these multi-turn interactions are not observable in single-turn test cases and found direct chat more informative. However, interviewees also remarked on the considerable workload of writing messages manually (I1), the difficulty of repeating conversations (I4), and the benefits of test cases over direct chats in terms of parallel reviewing (I2). I2 also commented that teachers would struggle to generate believable chats if they have less experience or teach humanities subjects whose content and patterns are diverse. 

\subsubsection{Teachers' mental model of review is based on student profiles, but they lack systematic approaches to organize and incorporate diverse types and granularities of student traits.} 
Interviewees created test cases and conversational patterns with specific student personas in mind and referred to them when explaining their rationale for test cases. For example, I4 recalled students with borderline intellectual functioning and tested if the PCA could provide digestible explanations and diagrams. However, interviewees tend to review PCAs on the fly without a systematic approach; interviewees mix different student personas (e.g., high and low knowledge, shy and active) in a single conversation instead of simulating each persona in a separate chat. I4 and I5 remarked that they had not conceived the separation, and single-persona conversations would have made the review more meaningful. I2 commented that creating student profiles first would have prepared her to organize more structural test cases. Interviewees also commented on the difficulty of describing students with varying levels within a trait (I4) and reflecting diverse traits in free-form writing (I1).

\subsection{Design Goals}
Based on the findings from the formative interview, we outline the design goals to help teachers efficiently review their PCAs' limitations against diverse students and improve their PCAs iteratively. The design goals are 1-to-1 mapped to each finding in \S\ref{sec:interview_findings} and aim to address teachers' needs and challenges.

\begin{enumerate}
    \item[\textbf{DG1.}] Support the reviewing of PCAs' adaptivity to students with varying knowledge levels, motivation factors, and psychosocial contexts.
    
    \item[\textbf{DG2.}] Offload the manual effort to generate multi-turn conversations for quick and iterative reviews in the PCA design process.
    
    \item[\textbf{DG3.}] Provide teachers with structures and interactions for authoring separate student profiles and organizing test cases.
\end{enumerate}

\section{System: \sysname{}}
We present \sysname{}, a web-based tool where teachers can build LLM-based PCAs and quickly review their coverage against simulated students with diverse knowledge levels, motivation factors, and psychosocial contexts before deploying the PCAs to actual students. We outline the user interfaces for creating PCAs, configuring simulated students of teachers' needs as test cases, and reviewing PCAs through automatically generated conversations between PCAs and simulated students. We also introduce our novel technical pipeline to simulate students behind the scenes.

\begin{figure*}[ht]
\includegraphics[width=\textwidth]{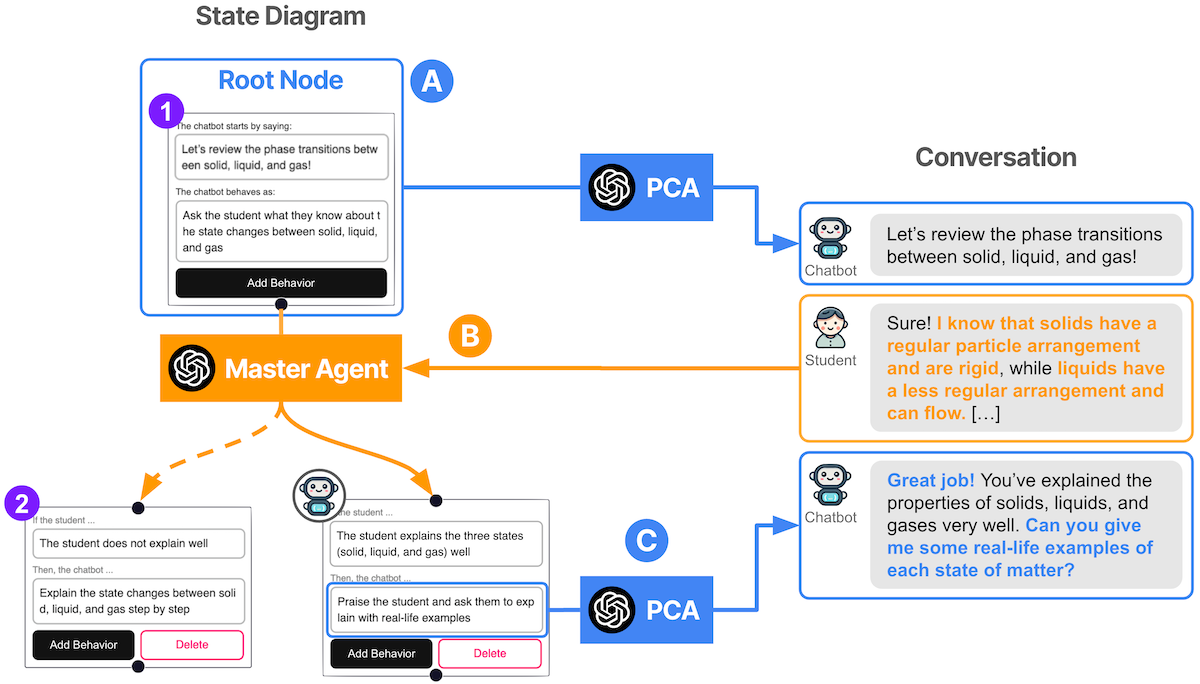}
    \caption{A PCA follows the dialogue flow defined in its state diagram. Nodes represent the PCA's utterance, and edges represent the potential response path of simulated students. The root node (A) contains the PCA's starting message and initial behavior. Based on a student's response, the master agent keeps the current state or changes the active node to one of the connected nodes (B). The next active node determines the PCA's subsequent response (C).}
    \label{fig:state_machine}
    \Description{Diagram showing how a state diagram and conversation are controlled by two LLM agents: PCA and Master Agent. On the left, the state machine diagram includes a root node and two child nodes. The root node represents an initial message generated by the PCA agent, which says, "Let's review the phase transition between solid, liquid, and gas." The student's message begins with "Sure! I know that [...]," followed by an explanation of the three states of matter. Based on the student's message, the master agent decides which child node to move to, selecting the node labeled "the student explains the three states well." This node contains an instruction: "Praise the student and ask them to explain with a real-life example." Following this, the final message generated by the PCA agent says, "Great job! Can you give me some real-life examples of each state of matter?"}
\end{figure*}

\begin{figure*}[ht]
\includegraphics[width=\textwidth]{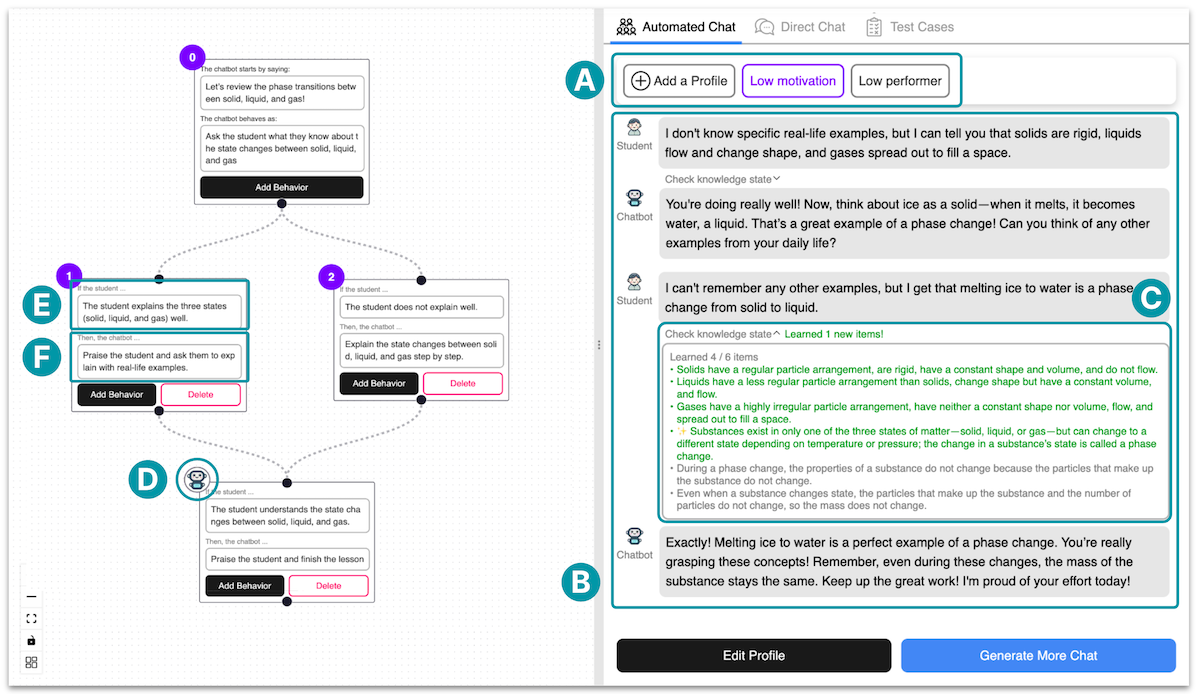}
    \caption{
    The \sysname{} interface. On the right, a teacher can add new student profiles (A) and review their auto-generated conversation (B). The teacher can also check the student's current knowledge stage at each utterance (C). On the left is the PCA creation interface with a state diagram. The robot icon shows the current state (i.e., active node) of the PCA at each turn (D). The PCA changes its behavior according to the conditions (E) and follows the instructions written on the currently active node (F).}
    \label{fig:interface_pca_creation}
    \Description{The interface of an automated chat showing a state machine for PCA development on the left and a chat interface on the right. On the right top is an "Add a profile" button for creating student profiles, with a conversation generated by a student named "Low motivation" displayed below it. Beneath the student's message, the "Check knowledge state" button has been pressed, highlighting the newly acquired knowledge elements in a different color. On the left side, the state diagram displays nodes with two text inputs each: the top text represents student behavior, such as "The student explains the three states well," and the bottom text represents instructions, such as "Praise the student and ask them to explain with real-life examples." At the left corner of the final node, a robot icon indicates the current active state.}
\end{figure*}

\subsection{PCA Creation Interface}
Teachers can build PCAs with a graph-like state machine representation (Fig.~\ref{fig:state_machine})~\cite{choi2021protochat, hedderich2024piece}. The state machine of a PCA starts with a root node that consists of the PCA's start message to students and the instruction it initially follows. For example, the PCA in Fig.~\ref{fig:state_machine} starts its conversation by saying: ``Let's review the phase transitions between solid, liquid, and gas!'' and asks questions about phase transitions to a student (Fig.~\ref{fig:state_machine} A) until the state changes to other nodes. The state changes to one of the connected nodes depending on whether or not the student answers the questions well (Fig.~\ref{fig:state_machine} B). When the state changes to either node, PCA receives a new instruction, described in the nodes, to behave accordingly (Fig.~\ref{fig:state_machine} C). The PCA is an LLM-based agent prompted conditionally with the state machine, whose state is determined by a master LLM agent. The master agent monitors the conversation between the PCA and a student and decides if the state should remain in the same node or transit to one of its child nodes. The prompts used to instruct the master agent and PCA are in Appendix~\ref{appendix:prompt_master_agent} and~\ref{appendix:prompt_pca_agent}.

\subsubsection{Authoring graph-based state machines}
\sysname{} provides a node-based interface to author the state machine of PCAs (Fig.~\ref{fig:interface_pca_creation} left). Teachers can drag to move nodes, zoom in, and pan the state diagram. They can add child nodes by clicking the ``Add Behavior'' button on the parent node. Teachers can also add directed edges between nodes to indicate the sequence of instructions PCAs should follow. In each node, teachers describe a student behavior for PCAs to react to (Fig.~\ref{fig:interface_pca_creation} E: ``if the student ...'') and instructions for PCAs to follow (Fig.~\ref{fig:interface_pca_creation} F: ``then, the chatbot ...''). Student behaviors are written in natural language, allowing teachers to cover a diverse range and granularity of cases, such as cases where students do not remember the term sublimation or ignore PCA's questions. Instructions can also take various forms, from prescribed explanations about sublimation to abstract ones, such as creating an intriguing question to elicit students' curiosity. To help teachers understand how the state machine works and debug it, \sysname{} visualizes a marker (Fig.~\ref{fig:interface_pca_creation} D) on the state machine diagram that shows the current state of PCA along conversations during reviews. The node-based interface helps teachers design and represent conversation flows that are adaptive to diverse cases.

\begin{figure*}[ht]
\includegraphics[width=\textwidth]{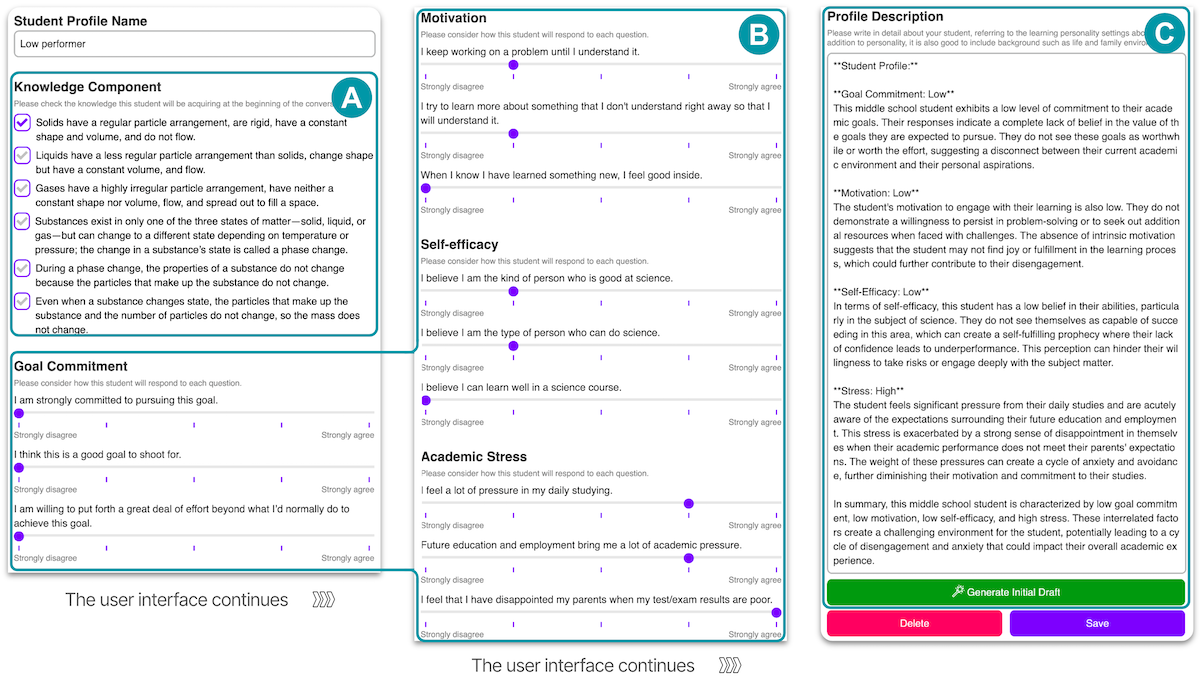}
    \caption{The interface to create a student profile. Teachers set the initial knowledge level of the student by check-marking the knowledge components to turn on at the beginning of a conversation (A). They also rate 5-point Likert scale questions to configure the four unique student traits (B). \sysname{} generates a (C) natural language student profile overview based on the information set from (B). Users can edit the system-generated description or add more contextual information about a student.}
    \label{fig:interface_profile_creation}
    \Description{Interface for creating a student profile in an automated chat. At the top, there are six knowledge components about phase transitions with checkbox inputs. After that, there are 12 slider inputs, each rated from 1 to 5, for rating four student traits with three questions per trait. Below the sliders is a "Generate initial draft" button. After clicking this button, a profile description is generated based on the selected knowledge components and student traits.}
\end{figure*}

\begin{figure*}[ht]
\includegraphics[width=\textwidth]{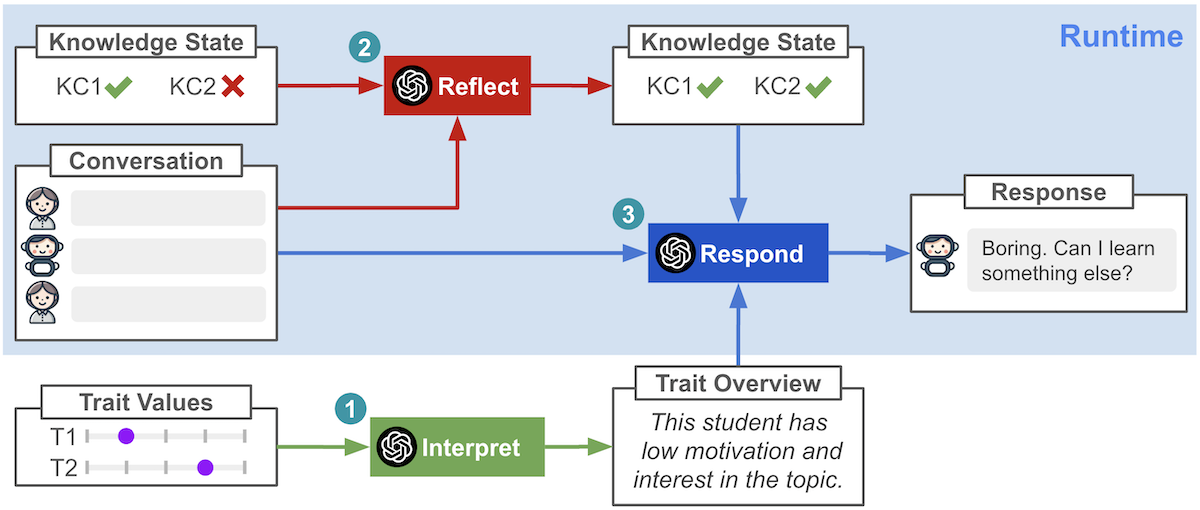}
    \caption{The \pipelinename{} pipeline. The pipeline interprets the student's trait values and creates a trait overview (1), and the previous conversation history is used to update the knowledge state through the reflect pipeline (2). Afterward, the Respond pipeline takes the conversation, updated knowledge state, and the trait overview to generate the response (3). The blue background is a runtime area where the components inside change throughout a conversation. The trait overview is created once before the runtime.}
    \label{fig:pipeline}
    \Description{A diagram illustrating a 3-step prompting pipeline for simulating student responses: Interpret, Reflect, and Respond. The Interpret step runs independently before runtime, creating a trait overview with the text, "This student has low motivation and interest in the topic," based on the student's trait values. During runtime, the previous knowledge state and conversation history flow into the Reflect step, where the knowledge state is updated. In the Respond step, the updated knowledge state, conversation history, and trait overview are used to generate the simulated student's response, shown as "Boring. Can I learn something else?"}
\end{figure*}

\subsection{PCA Review Interface}
Teachers can review the robustness of their PCAs by testing different edge cases with three methods---direct chat, single-turn test cases, and automated chat. The user interface for direct chat and test cases are identical to the ones used in the formative study (Fig.~\ref{fig:formative_system}); teachers can either directly talk to their PCAs over multi-turn or test multiple pre-defined input messages at once and observe how PCAs respond to each. The last and our novel method, review through automated chats, involves two steps---creating student profiles and observing simulated conversations.

\subsubsection{Templated student profile creation}
Teachers should first define what types of students they review against. \sysname{} helps teachers externalize and develop their evaluation space with templated student profiles. Our interface (Fig.~\ref{fig:interface_profile_creation}) provides knowledge components and student trait inventories to help teachers recognize possible combinations and granularities of different knowledge levels and traits and organize them effectively (DG3). When creating each student profile, teachers can specify the student's initial knowledge by check-marking knowledge components (Fig.~\ref{fig:interface_profile_creation} A) and configure the student's personality by rating the trait inventories on a 5-point Likert scale (Fig.~\ref{fig:interface_profile_creation} B). \sysname{} then generates a natural language description of the student, which teachers can freely edit to correct or add more contextual information about the student (Fig.~\ref{fig:interface_profile_creation} C). This description, namely \textbf{trait overview}, is passed to our simulation pipeline.

Once teachers have created a pool of student profiles to review against, they can leverage it over their iterative PCA design process, like how single-turn test cases are efficient for repeated reviews. We decided to let teachers configure their student pools instead of automatically providing all possible student profiles because it is time-consuming for teachers to check student profiles who might not even exist in their classes. 

\sysname{} populates knowledge components pre-defined in textbooks and curricula. Teachers can also add custom (e.g., more granular) knowledge components. For the trait inventories, we chose the top three statements from existing inventories~\cite{sun2011educational, klein2001assessment, gottfried1985academic, may2009mathematics} based on their correlation to student performance (see Appendix~\ref{appendix:kc_traits}). We present three statements for each trait, considering the efficiency and preciseness in authoring student profiles, heuristically decided from our iterative system design.

\subsubsection{Automated chat}
Teachers then select one of the student profiles to generate a lesson conversation between the profile's simulated student and their PCAs (Fig.~\ref{fig:interface_pca_creation} A). PCAs start conversations, and the state marker on the state diagram transits in real-time throughout the conversation. Simulated students initially show unawareness as prescribed by their knowledge states in profiles and acquire knowledge from PCAs in mock conversations. Simulated students also actively ask questions, show indifference, or exhibit passive learning attitudes according to their student traits. \sysname{} generates six messages (i.e., three turns) between PCAs and simulated students at a time, and teachers can keep generating further conversation by clicking the ``Generate Conversation'' button. When teachers change the state machine diagram, \sysname{} prompts teachers to re-generate conversations from the beginning. Teachers can use automated chats to quickly review different PCA designs on the same students without manually typing messages (DG2). When teachers find corner cases that their PCA design did not cover, they can add a node that describes the case and appropriate instruction for PCAs. For example, with the state machine in Fig.~\ref{fig:state_machine}, teachers may find the PCA stuck in the root state when it chats with a simulated student who asks questions. To handle the case, teachers can add a node that reacts to students' questions and instruct PCA to answer them.

\subsection{\pipelinename{}}
We propose a \pipelinename{} LLM pipeline that simulates conversations with specific student profiles. Our pipeline design is inspired by and extended from Jin et al.'s Reflect-Respond pipeline~\cite{jin2024teach}; we added a personalization component that prompts LLMs to incorporate prescribed student traits into simulated students (DG1).

Reflect-Respond is an LLM-driven pipeline that simulates knowledge-learning~\cite{jin2024teach}. It takes a simulated student's current knowledge state and conversation history as inputs (Fig.~\ref{fig:pipeline}). A knowledge state is a list of knowledge components that are either acquired or not acquired. The state dynamically changes throughout conversations to mimic knowledge acquisition. To generate a simulated student's response, inputs pass through the \textit{Reflect} and \textit{Respond} steps. \textit{Reflect} updates the knowledge state by activating relevant components, while \textit{Respond} produces a likely reply based on the updated state and conversation history.

Our pipeline personalizes Reflect-Respond by giving an LLM additional instruction in the \textit{Respond} step. Before the runtime of Reflect-Respond, \textit{Interpret} step first translates trait scores into a \textbf{trait overview} that contains a comprehensive summary and reasoning of how the student should behave (Fig.~\ref{fig:pipeline} Step 1). Once teachers edit and confirm the overview through the interface (Fig.~\ref{fig:interface_profile_creation} C), it is passed to the \textit{Respond} step so that the LLM takes the student traits into account in addition to the conversational context and knowledge state. We added the \textit{Interpret} step because it produces student profiles that allow teachers to edit flexibly and prompt LLMs to reflect on student traits more cohesively (i.e., chain of thought~\cite{wei2022chain}). The prompts for \textit{Interpret}, \textit{Reflect}, and \textit{Respond} are available in Appendix~\ref{appendix:prompt_profile},~\ref{appendix:prompt_reflect}, and~\ref{appendix:prompt_respond}. 

We took an LLM-driven approach to personalize and implement the Reflect-Respond pipeline. We considered adopting student modeling methods that rely on more predictable and grounded Markov models~\cite{tadayon2020predicting, maqsood2022modeling}. Still, we decided to use a fully LLM-driven approach because we also target extracurricular teaching scenarios where large datasets to build Markov models may not be available.
\section{Evaluation}
We evaluated the alignment of \pipelinename{} to teachers' perception of simulated students and the efficacy of \sysname{} for helping teachers review PCAs against diverse student profiles. Our evaluation explores the following research questions:

\begin{enumerate}
    \item[\textbf{RQ2:}] How accurately does the \pipelinename{} pipeline simulate a student's knowledge level and traits expected by teachers?
    
    \item[\textbf{RQ3:}] How do simulated students and automated chats, compared to direct chats and test cases, help teachers review PCAs?
\end{enumerate}

The evaluation was twofold. To investigate RQ2, we created nine simulated students of diversely sampled knowledge and trait configurations and asked 10 teachers to predict their configurations through direct chats and pre-generated conversations. To answer RQ3, we ran a between-subjects user study with 30 teachers and observed how the student profile template and simulated students helped the design and reviewing of PCAs. We received approval for our evaluation study design from our institutional review board.

\subsection{Technical Evaluation}
Under controlled settings, we evaluated how well the behavior of a simulated student instructed by our pipeline aligns with teachers' expectations of the student regarding knowledge level, motivation, and psychosocial contexts (RQ2).

\subsubsection{Evaluators}
We recruited ten K-12 science teachers as evaluators through online teacher communities. The evaluators had experience teaching 25$\pm$9.2-sized ($\mu\pm\sigma$) classes (min: 8, max: 33) for 4.5$\pm$4.2 years (min: 0.5, max: 15). As compensation, evaluators received KRW 50,000 (USD 38).

\subsubsection{Baseline Pipeline}
We created a baseline pipeline to explore how the \textit{Interpret} step affects the alignment gap. The \textit{\textbf{Baseline}} pipeline directly takes raw student traits in its \textit{Respond} step without the \textit{Interpret} step. By comparing \textit{Baseline} with \textit{\textbf{Ours}} (i.e., \pipelinename{}), we aimed to investigate if explanation-rich trait overviews help an LLM reduce the gap between simulated students and teachers' expectations. Pipelines were powered by GPT-4o-mini, with the temperature set to zero for consistent output. The prompt used for \textit{Baseline} is available in Appendix~\ref{appendix:baseline_prompt}.

\subsubsection{Setup}
The phase transition between solid, liquid, and gas was the learning topic of our setup. We chose phase transition because it has varying complexities of knowledge components and applicable pedagogies. Simulated students could initially know and learn six knowledge components of varying complexity (see Appendix~\ref{appendix:kc_traits}); the first three components describe the nature of three phases, and the latter three are about invariant properties in phase transition with reasoning. The knowledge components were from middle school science textbooks and curricula qualified by the Korean Ministry of Education. 

\begin{figure*}[ht]
\includegraphics[width=\textwidth]{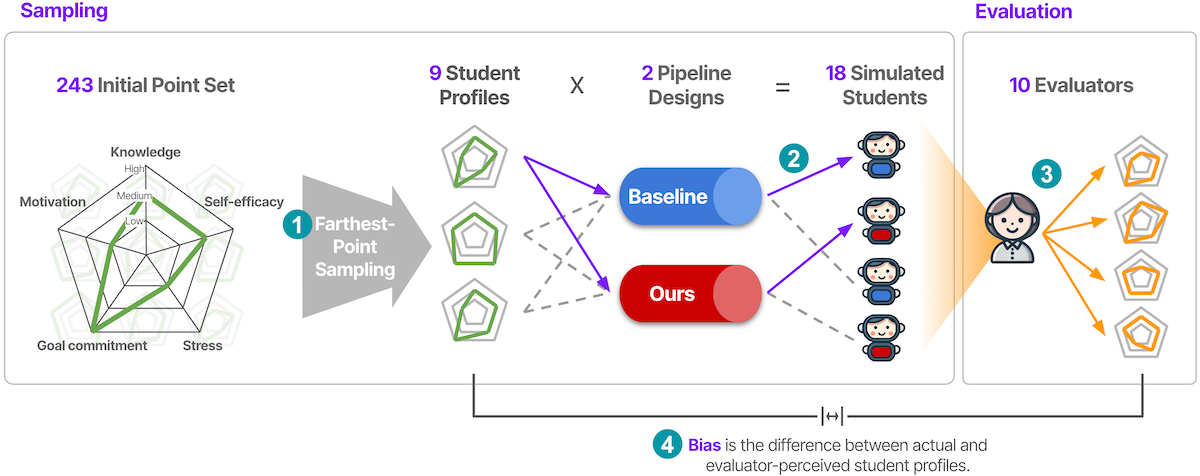}
    \caption{A summary of our technical evaluation. From the 243 possible combinations of intensities (3 levels (high/medium/low) for each of the five characteristics), we used farthest-point sampling (1) to sample nine unique student profiles. Then, we ran each of the nine student profiles in the \textit{Baseline} and \pipelinename{} pipeline, which resulted in 18 simulated students (2). A total of 10 evaluators were recruited to predict the student profiles given conversation histories in a blind condition (3). We then measured bias between generated student profiles and evaluators' predicted student profiles (4).}
    \label{fig:tech_eval_process}
    \Description{Diagram summarizing key steps of the technical evaluation process. Overall, five traits (knowledge level, motivation, goal commitment, stress, and self-efficacy) are visualized as radar charts, representing student profiles. First, from an initial set of 243 points, 9 student profiles were derived using farthest-point sampling. These profiles were then used to generate 18 simulated students through two different pipeline designs: Baseline and Ours. Finally, 10 evaluators inferred the profiles of the simulated students under blind conditions. Bias is depicted as the difference between the radar charts representing the actual profiles from the pipeline and the evaluator-perceived profiles.}
\end{figure*}

We prepared 18 simulated students for the evaluation (see Fig.~\ref{fig:tech_eval_process}). We first chose nine student profiles through the farthest-point sampling~\cite{qi2017pointnet++}, where the point set was 243 possible combinations of different levels of knowledge and student traits to ensure the coverage and diversity of samples. Each student profile was instantiated into two simulated students instructed by \textit{Baseline} and \textit{Ours}.

\subsubsection{Procedure}
We first explained the research background to the evaluators. The evaluators then reviewed 18 simulated students independently in a randomized order. To reduce fatigue from conversing with simulated students manually, we provided two pre-prepared dialogues---interview and lesson dialogues. In interview dialogues, simulated students sequentially responded to six quizzes about phase transition and ten questions about their student traits (Fig.~\ref{fig:interview_dialogue}). In lesson dialogues, simulated students received 12 instructional messages dynamically generated by an LLM tutor prompted to teach phase transitions (Fig.~\ref{fig:lesson_dialogue}). Lesson dialogues show more natural teacher-student conversations in which teachers speak adaptively to students. Evaluators could also converse with simulated students directly if they wanted. Nine evaluators used direct chats at least once; they conversed with 5$\pm$4.5 students and exchanged 8$\pm$8.3 messages on average. 

We gave evaluators a list of six knowledge components and three 5-point Likert scale inventory items for each student trait; they predicted each simulated student's initial knowledge state, intensity level of the four student traits, and believability. The sampled student profiles, trait overviews, knowledge components, and inventory items used are available in Appendix~\ref{appendix:sample_parameters},~\ref{appendix:profile_description}, and~\ref{appendix:kc_traits}.

\subsubsection{Measures}
We measured the alignment between simulated students' behaviors and teachers' expectations of them in two aspects---bias and believability. The bias is the gap between the teacher-perceived and system-configured student profiles. A smaller bias would indicate that our pipeline simulates student behaviors closer to what teachers anticipate. Believability~\cite{park2023generative} is the perceived authenticity of simulated students regarding their response content and behavior patterns. We measured the bias and believability of each sampled student profile independently and analyzed the overall trend.

Evaluators' marking on knowledge components was binary (i.e., whether a simulated student possesses this knowledge), and their rating on the four student traits was a scalar ranging from three to fifteen, summing 5-point Likert scale scores from three inventory items as originally designed~\cite{sun2011educational, klein2001assessment, gottfried1985academic, may2009mathematics}. We used the two-sided Mann-Whitney U test per simulated student pairs to compare \textit{Baseline} and \textit{Ours}. We report the following measures:

\begin{itemize}
    \item \textbf{Knowledge Bias} (\% error). We quantified the bias on knowledge level as the percentage of incorrectly predicted knowledge components. We report the average and median across the evaluators.

    \item \textbf{Student Trait Bias} (0-12 difference). We calculated the mean absolute error between the evaluators' Likert score and the configured value for each student trait. We report the average and median across the evaluators.

    \item \textbf{Believability} (1-5 points). We directly compared evaluators' ratings on the three statements about how authentic simulated behavioral and conversational responses are and how valuable simulated students are for teaching preparation (Fig.~\ref{fig:tech_eval_result_believability}).
\end{itemize}

\begin{figure*}[ht]
    \input{table/interview_dialogue}
    \caption{An interview dialogue of a simulated student powered by \pipelinename{} with the following student profile: medium knowledge level, high goal commitment, medium motivation, low self-efficacy, and low stress. The dialogue was designed to reveal the characteristics of the student.}
    \label{fig:interview_dialogue}
    \Description{An interview conversation between an interviewer and a simulated student. The dialogue focuses on reviewing the characteristics of solids and liquids, changes of state, and personal science study goals. The student expresses nervousness but is willing to try a quiz and shares aspirations and concerns about persistence in science studies.}
\end{figure*}

\begin{figure*}[ht]
    \input{table/lesson_dialogue}
    \caption{A lesson dialogue of the teacher chatbot and a simulated student powered by \pipelinename{} with the following student profile: medium knowledge level, high goal commitment, medium motivation, low self-efficacy, and low stress. The dialogue was designed to simulate a review session.}
    \label{fig:lesson_dialogue}
    \Description{An interview conversation between an LLM tutor and a simulated student. They review the particle arrangements in solids, liquids, and gases, as well as particle motion and state changes. The tutor reassures the student and explains concepts not previously covered, while the student expresses uncertainty but finds the topics interesting.}
\end{figure*}

\subsection{Technical Evaluation Result}
We report the descriptive statistics on the bias and believability of \pipelinename{} (\textit{Ours}) and validate its design by comparing it with \textit{Baseline}. Our results collectively show that \pipelinename{} can instruct an LLM to simulate a student's behavior of a specific knowledge state and traits precisely. 

\subsubsection{The knowledge bias was small (median: 5\%)}
The gap between the configured and evaluator-perceived knowledge states was small (the last row of Table~\ref{table:tech_eval_result_knowledge}). Among the nine student profiles, evaluators unanimously and correctly identified the knowledge components of four profiles. The average accuracy across profiles was 93\%, where the minimum was 78\%. Profiles 4 and 7 achieved the lowest accuracy; evaluators underrated Profile 4 and overrated Profile 7. Student profile 4 describes a learner who knows all knowledge components but exhibits low confidence and interest. The corresponding simulated student tended to respond to the tutor's questions half-heartedly. We speculate that this behavior might have confused evaluators to think the student was unaware of some of the knowledge components. Student profile 7 was a learner who knew only half of the knowledge but had high self-efficacy. Its confident response might have deluded evaluators that it knows more.

\subsubsection{The trait bias was small (median: 1.3 out of 12)}
The gap between the configured and perceived levels of student traits was also small (Fig.~\ref{fig:tech_eval_result_trait}). The mean bias was 1.9, and the minimum and maximum were 0.4 and 4.9, respectively. Considering that we summed the bias from three 5-point scale questions for each trait, teachers can precisely set their simulated students within less than $\pm1$ point error on each Likert scale input in our profile generation interface (Fig.~\ref{fig:interface_profile_creation} B). The average variance between the perceived traits was also small ($\sigma^2=0.61$), possibly indicating that simulated students manifested characteristics unique to their traits and led to a high agreement among teachers' perceptions. Nevertheless, Profiles 3, 4, and 9 showed biases above four on the goal commitment trait. All of these student profiles had contrasting goal commitment and motivation ratings; for instance, the goal commitment rating of Profile 3 was low, while the motivation rating was high. We contemplate that since these two traits often correlate and go together~\cite{sue2002goal, mikami2017relationships}, evaluators might have misunderstood the motivational behaviors of simulated students as goal-related patterns.

\subsubsection{Simulated students were believable (median: 3.5 out of 5)}
Evaluators reported that simulated students behave as naturally as real students and are helpful for teacher training (Fig.~\ref{fig:tech_eval_result_believability}). The average scores for each question (i.e., B1, B2, and B3) were $3.6\pm0.4$, $3.5\pm0.3$, and $3.4\pm0.3$, respectively. The variance in the B1 scores was high in some of the profiles. For instance, the variance was 2.1 (min: 1, max: 5) for Profile 2, which describes a student with zero knowledge and the lowest goal commitment, motivation, and self-efficacy. Since the simulated student knew nothing, it repeatedly said ``I do not know'' in its interview and lesson dialogues as instructed by its prompt (Appendix~\ref{appendix:prompt_respond}). Evaluators had different opinions on this behavior; low raters felt the repetitive messages were unnatural and artificial, while high raters thought unmotivated students tended to give short and sloppy answers in actual classes. B3 scores showed a similar trend and a high correlation to B1 scores (Pearson's $r$=0.96).

\subsubsection{The Interpret step increased believability significantly}
\begin{figure*}[ht]
    \includegraphics[width=\textwidth]{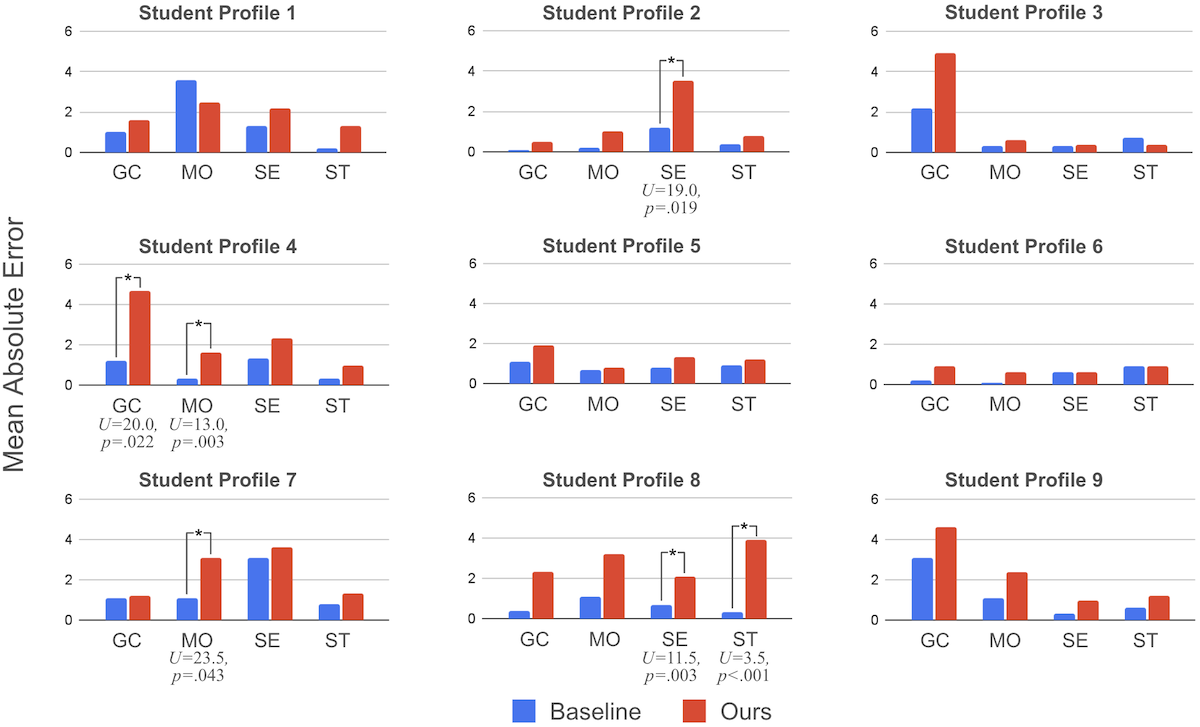}
    \caption{The bias in four student traits: goal commitment (GC), motivation (MO), self-efficacy (SE), and stress (ST). The asterisk (*) indicates statistical significance ($p$<.05) between conditions. For specific student profiles, please refer to Appendix \ref{appendix:sample_parameters}.}
    \label{fig:tech_eval_result_trait}
    \Description{Bar graphs comparing the mean absolute error across two conditions, Baseline and Ours, for nine student profiles. The x-axis represents four traits (goal commitment, motivation, self-efficacy, and stress), and the y-axis ranges from 0 to 6. For Student Profile 2, the self-efficacy trait shows a significant difference in error, with a U-value of 19.0 and a p-value of 0.019. For Student Profile 4, both goal commitment and motivation traits show significant differences in error, with U-values of 20.0 and 13.0 and p-values of 0.022 and 0.003, respectively. For Student Profile 7, the motivation trait shows a significant difference in error, with a U-value of 23.5 and a p-value of 0.043. For Student Profile 8, both self-efficacy and stress traits show significant differences in error, with U-values of 11.5 and 3.5 and p-values of 0.003 and less than 0.001, respectively.}
\end{figure*}

\begin{figure*}[ht]
    \includegraphics[width=\textwidth]{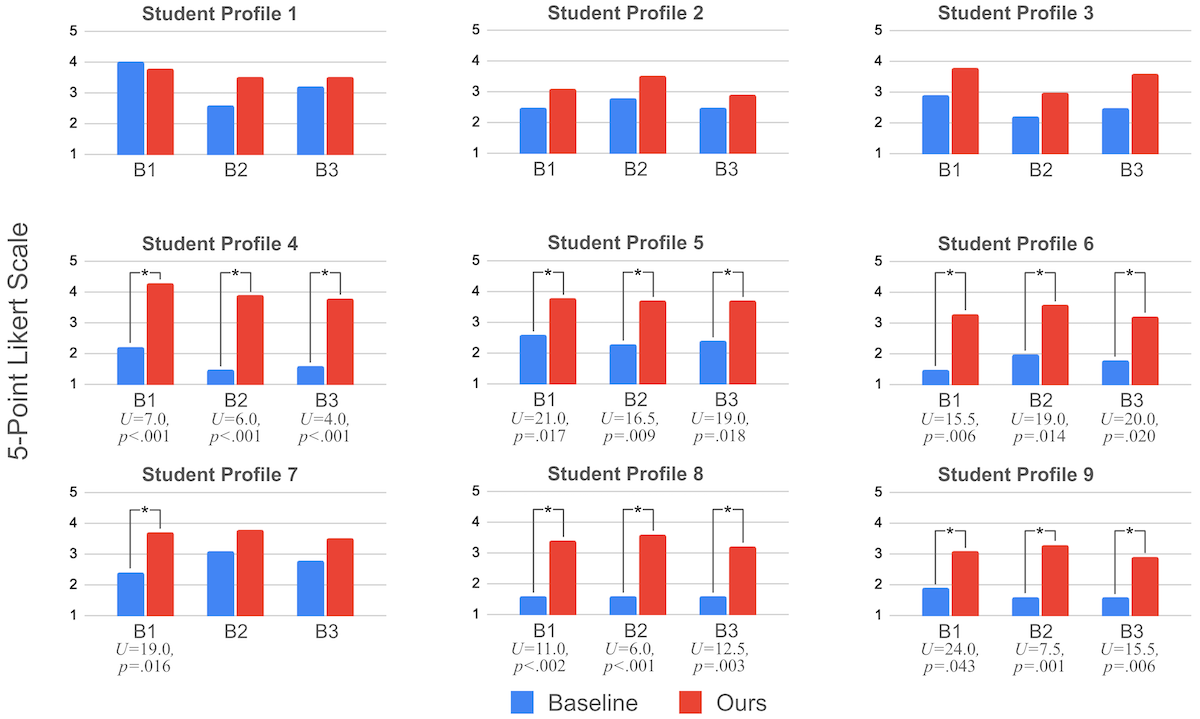}
    \caption{Result of the believability measured in 5-point Likert scale (1: Strongly disagree, 5: Strongly agree) with three questions. B1: This student naturally responds (e.g., explain, question, ignore) to the teacher's questions or instructions. B2: This student uses language and speaking style that a real student would use. B3: This student looks real and is useful as a chatbot for teacher training. The asterisk (*) indicates statistical ($p$<.05) significance between conditions. For specific student profiles, please refer to Appendix \ref{appendix:sample_parameters}.}
    \label{fig:tech_eval_result_believability}
    \Description{Bar graphs comparing the believability ratings measured on a 5-point Likert scale across two conditions, Baseline and Ours, for nine student profiles. The x-axis represents three questions, and the y-axis ranges from 1 to 5. For Student Profiles 4, 5, 6, 8, and 9, there are statistically significant differences in scores for all three questions, with p-values all below 0.05. For Student Profile 7, the score for the first question shows a significant statistical difference, with a U-value of 19.0 and a p-value of 0.016.}
\end{figure*}

Our ablation study showed the tradeoff relationship between the bias and believability in our pipeline design. The \textit{Baseline} pipeline showed minimal knowledge and trait bias compared to \textit{Ours} (Table~\ref{table:tech_eval_result_knowledge} and Fig.~\ref{fig:tech_eval_result_trait}). Bias was minimal because \textit{Baseline} students often revealed the raw trait values in the system prompt when responding to questions (e.g., ``I have a low motivation'' and ``I strongly agree.'') However, these frank responses resulted in a statistically significant decrease in the believability of simulated students (Fig.~\ref{fig:tech_eval_result_believability}). Evaluators felt artificiality towards the dry and repeated responses and perceived them as detrimental to being a pedagogy tester (B3).
On the other hand, \textit{Ours} students were better at incorporating multiple traits into responses. For example, Profile 5 is a student who has high goal commitment and stress levels at the same time. While \textit{Baseline} generated ``Thank you! But, I am stressed about my daily study.'' for a tutor's encouragement, \textit{Ours} creates a multifaceted response: ``Thank you! I am a bit stressed about my daily study, but I am trying hard.'' The \textit{Interpret} step can balance the tradeoff between bias and believability by prompting LLMs to analyze student profiles more comprehensively and generate more believable behaviors.

\begin{table*}[ht]
\centering
\begin{tabular}{lccccccccccc}
\toprule
 & \multicolumn{11}{c}{Student Profiles}                                                     \\ \cline{2-12} 
 & 1 & 2 & 3 & 4 & 5 & 6 & 7 & 8 & 9 & \multicolumn{1}{l}{Mean} & \multicolumn{1}{l}{Median} \\ \hline
\multicolumn{1}{c}{Baseline} & 8.3$\pm$20.4  & 0.0$\pm$0.0  & 0.0$\pm$0.0 & 13.3$\pm$5.2  & 1.7$\pm$4.1 & 10.0$\pm$0.0 & 6.7$\pm$5.2   & 0.0$\pm$0.0 & 0.0$\pm$0.0 & 4.4 & 1.7 \\
\multicolumn{1}{c}{Ours}     & 8.3$\pm$.11.7 & 6.7$\pm$16.3 & 5.0$\pm$5.5 & 21.7$\pm$20.4 & 0.0$\pm$0.0 & 0.0$\pm$.0.0 & 21.7$\pm$18.3 & 0.0$\pm$0.0 & 0.0$\pm$0.0 & 7.0 & 5.0 \\ 
\bottomrule
\end{tabular}%
\caption{The knowledge bias of each student profile. The bias was overall small, with an average of 7\%, with Profile 4 and 7 having the largest bias.}
\Description{Table comparing knowledge bias for nine student profiles across two conditions: baseline and ours. Each row lists the mean and standard deviation values, with the last two columns showing the mean and median bias.}
\label{table:tech_eval_result_knowledge}
\end{table*}

\subsection{User Study}
We ran a user study with 30 K-12 science teachers to explore how templated student profile creation and automated chats affect the PCA design process (RQ3). We designed a between-subjects study in which each participant created a PCA under one of the three conditions—\textit{Baseline}, \textit{Autochat}, and \textit{Knowledge}. In \textbf{Baseline}, participants used a version of \sysname{} without the automated chat feature; participants could access direct chat and single-turn test cases only. In \textbf{Autochat}, participants used \sysname{} with all features available; they could generate student profiles with our template interface and use automated chats, direct chats, and test cases. In \textbf{Knowledge}, participants used another version of \sysname{} where they could use all features but configure only the knowledge level of simulated students (i.e., no student traits and trait overview); this is analogous to using simulated students powered by the original Reflect-Respond pipeline. 

By comparing the three conditions, we investigated the effect of having simulated students on PCA review (\textit{Baseline} vs. \textit{Autochat}) and how simulating student traits beyond their knowledge level affect the depth and breadth of the design process (\textit{Autochat} vs. \textit{Knowledge}). The \textit{Knowledge} condition is the baseline for the automated chat feature. By looking into this condition, we investigate if the existing simulated student pipeline (i.e., Reflect-Respond) is enough to elicit improved test coverage and how \pipelinename{} can improve it further.

\subsubsection{Participants}
We recruited 36 teachers through online teacher communities in Korea and randomly assigned them to one of the conditions. Participants had varying teaching periods (3.3$\pm$4.7 years) and class sizes (13$\pm$12 students). Thirteen participants are currently teaching at public schools. According to our pre-task survey (Appendix~\ref{appendix:pre_task_questions}), all participants had experience using chatbots and ChatGPT. They responded that they were interested in using AI (e.g., image generation AI and ChatGPT) in their classes. More than half of the participants reported they were knowledgeable about the chatbot design process, and five of them actually had experience making chatbots. There was no statistical difference in participants' teaching experience, openness to AI technology, and knowledge about chatbot design among the conditions. Study sessions took place for 1.5 hours, and participants received KRW 50,000 (USD 38) as compensation. 

We randomly assigned ten participants to each condition, and the study was run asynchronously, considering participants' geographical diversity and daytime teaching positions. We also conducted additional sessions with six teachers in \textit{Autochat} condition to complement our asynchronous study design by observing how teachers interact with \sysname{} directly through Zoom screen sharing. We monitored the whole session and asked questions during and after they created PCAs. We excluded these six participants from our comparative analysis due to our intervention within the sessions. We only report their comments.

\subsubsection{Procedure and Materials}
After submitting informed consent, the participants received an online link to our system and completed a given task in their available time, following the instructions on the website. Participants first read an introduction about the research background and the purpose of this study and watched a 7-minute tutorial video about the features in \sysname{}. Participants could revisit the tutorial materials anytime during the study.

We asked participants to use \sysname{} to create a PCA that can teach ``the phase transitions between solid, liquid, and gas'' to students of as diverse knowledge levels and student traits as possible. Participants then used \sysname{} in one of the \textit{Baseline}, \textit{Autochat}, and \textit{Knowledge} conditions to design their PCAs for 30-60 minutes; participants spent 50$\pm$15 minutes on average. All participants received a list of knowledge components for the topic and explanations of the four student traits to ensure consistency and prevent bias in information exposure. We encouraged participants to consider them throughout the design process. After completing their PCA design, participants rated their task load. Participants then revisited their direct chats, test cases, simulated students, and state diagrams to report the student profiles they had considered in a predefined template (Fig.~\ref{fig:profile_collection_ui}). The study finished with a post-task survey asking about their PCA design experience. The study procedure is summarized in Table~\ref{tab:study_procedure}.

\subsubsection{Materials and Setup}
Participants received the six knowledge components used in our technical evaluation. We also gave participants an initial state diagram to help them start their PCA design. The knowledge components, initial state diagram, and survey questions are available in Appendix~\ref{appendix:kc_traits},~\ref{appendix:pre_task_questions},~\ref{appendix:post_task_questions}, and~\ref{appendix:start_state_diagram}.

We also made a few modifications to our pipeline setup. Our technical evaluation revealed that repeated responses critically undermine simulated students' perceived believability and usefulness. To prevent repeated responses and improve the efficacy of the automated chat, we set the temperature of the \textit{Respond} step to 1.0 and added a short instruction on repetition at the end of the prompt (Appendix~\ref{appendix:prompt_respond} red text). The prompt and temperature for other pipeline components were the same as the technical evaluation.

\subsubsection{Measures}
We looked into how \sysname{} affects the PCA design process as a review tool. An ideal review tool would help users reduce manual task loads, explore extensive evaluation space, and create quality artifacts. We evaluated each aspect with the following measures. Since we had a small sample size for each condition (n=10) and it was hard to assume the normality, we statistically compared the measures between the conditions through the Kruskal-Wallis test. We conducted Dunn's test for post hoc analysis.

\begin{figure*}[ht]
\includegraphics[width=\textwidth]{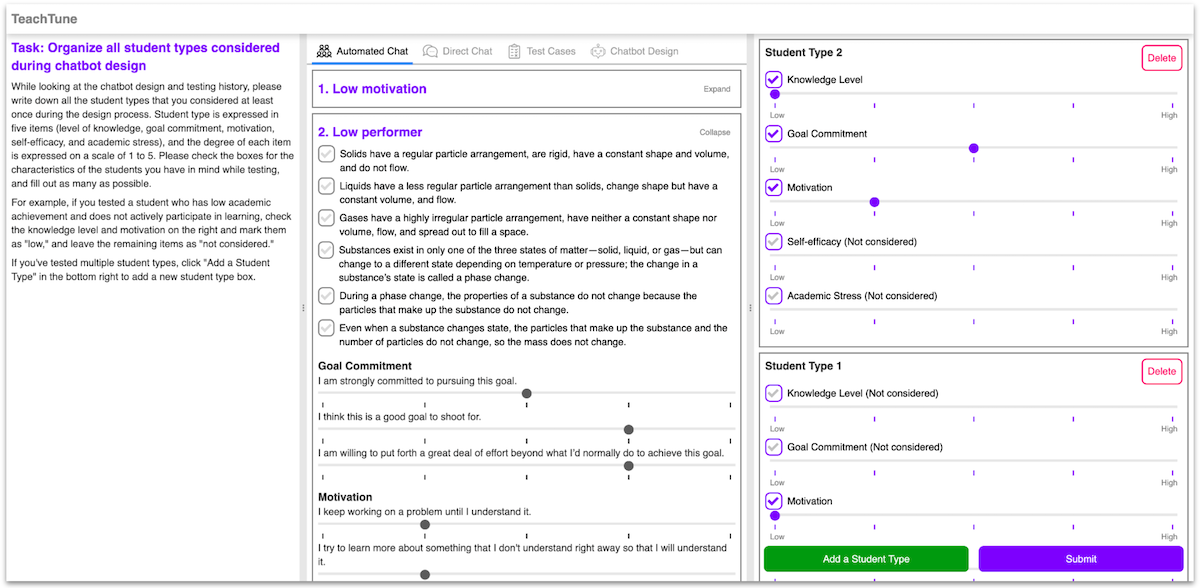}
    \caption{The profile collection UI used in Step 4: student profile reporting. The participants were instructed to report the types of students they considered in their chatbot design on the right in the unit of profiles containing knowledge and traits. In this process, they had access to history, including their automated chats, direct chats, and test cases, as well as the designed PCAs.}
    \label{fig:profile_collection_ui}
    \Description{An interface for collecting student profiles with three panels. The left panel provides instructions for the task. The middle panel shows the history, featuring two student profiles labeled "Low motivation" and "Low performer," which are used for automated chat methods. The right panel is for reporting student profiles. It includes five sliders that can be checked or unchecked, representing traits such as knowledge level, goal commitment, motivation, self-efficacy, and academic stress. In the displayed profile, knowledge level, goal commitment, and motivation are set to 1, 3, and 2 points, respectively, while self-efficacy and academic stress are unchecked, indicating they were not considered. At the bottom of the right panel, two buttons are displayed: an "Add a student type" button for adding additional profiles and a "Submit" button for final submission.}
\end{figure*}

\begin{itemize}
    \item \textbf{Task load} (1-7 points). Participants responded to the 7-point scale NASA Task Load Index~\cite{hart1988development} right after building their PCAs (Table~\ref{tab:study_procedure} Step 3). We modified the scale to seven to make it consistent with other scale-based questionnaires. Participants answered two NASA TLX forms, each asking about the task load on PCA creation and PCA review tasks, respectively.
    
    \item \textbf{Coverage.} We asked participants to report the student profiles they have considered in their design process (Table~\ref{tab:study_procedure} Step 4). We gave a template where participants could indicate each of the knowledge levels and four student traits of a student profile into five levels (1: very low, 5: very high). Participants could access their usage logs of direct chats, single-turn test cases, automated chats, and state diagrams to recall all the student profiles covered in their design process (Fig.~\ref{fig:profile_collection_ui}). We define \textit{coverage} as the number of unique student profiles characterized by the combinations of levels. We focused only on the diversity of knowledge levels and four traits to compare the conditions consistently. We chose self-reporting because system usage logs cannot capture intended student profiles in \textit{Baseline} and \textit{Knowledge}.

    \item \textbf{Quality} (3-21 points per trait). Although our design goals center around improving the coverage of student profiles, we also measured the quality of created PCAs. This was to check the effect of coverage on the final PCA design. We asked two external experts to rate the quality of the PCAs generated by the participants. Both experts were faculty members with a PhD in educational technologies and learning science and have researched AI tutors and pedagogies for ten years. The evaluators independently assessed 30 PCAs by conversing with them and analyzing their state machine diagrams. Evaluators exchanged a median of 28$\pm$10 and 45$\pm$20 messages per PCA. We instructed the evaluators to rate the heuristic usability of PCAs~\cite{langevin2021heuristic} and their coverage for knowledge levels and student traits (Appendix~\ref{appendix:chatbot_quality_criteria}). The usability and coverage ratings were composed of three 7-point scale sub-items, and we summed them up for analysis. Evaluators exchanged their test logs and ratings for the first ten chatbots to reach a consensus on the criteria. If the evaluators rated a PCA more than 3 points apart, they rated the PCA again independently. We report their mean rating after conflict resolution.

    \item \textbf{Post-task Survey.} We asked participants about the usefulness of each PCA review method and satisfaction on a 7-point Likert scale (Table~\ref{tab:study_procedure} Step 5). We also collected free-form comments from participants about their rationale for ratings (Appendix~\ref{appendix:post_task_questions}).
\end{itemize}

\begin{table}[ht]
\begin{tabular}{cc}
\toprule
\textbf{Step (min.)} & \textbf{Activity}                                      \\ \hline
1 (10)               & \makecell{Introduction on research background \\ and user interface} \\
2 (60)               & PCA design                                             \\
3 (5)                & Task load measurement                                  \\
4 (10)               & Student profile reporting                              \\
5 (5)                & Post-task survey                                \\ 
\bottomrule
\end{tabular}%
\caption{The study procedure. A single study session took around 90 minutes in total, and the participants were given 60 minutes for the PCA design.}
\Description{Table outlining the steps and time allocation for each activity. Step 1 (10 minutes) involves an introduction to the research background and user interface. Step 2 (60 minutes) is for PCA design. Step 3 (5 minutes) covers task load measurement. Step 4 (10 minutes) involves student profile reporting. Step 5 (5 minutes) is for a post-task questionnaire.}
\label{tab:study_procedure}
\end{table}

\subsection{User Study Result}
Participants created PCAs with 15$\pm$6 nodes and 21$\pm$10 edges in their state diagram on average. We outline the significant findings from the user study along with quantitative measures, participants' comments, and system usage logs. Participants are labeld with B[1-10] for \textit{Baseline}, A[1-10] for \textit{Autochat}, K[1-10] for \textit{Knowledge}, and O[1-6] for the teachers we directly observed.

\subsubsection{Autochat resulted in a lower physical and temporal task load}
\begin{figure*}[ht]
\includegraphics[width=\textwidth]{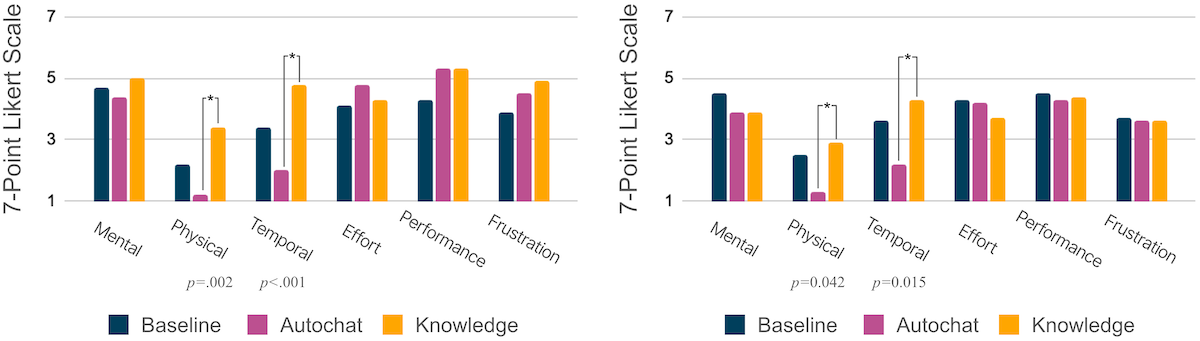}
    \caption{NASA-TLX survey results for PCA creation task (left) and PCA review task (right). The asterisk (*) indicates statistical significance ($p$<0.05) between conditions.}
    \label{fig:result_task_load}
    \Description{Two sets of bar graphs comparing the task load on a 7-point Likert scale across three conditions: Baseline, Autochat, and Knowledge. The x-axis represents six factors: mental, physical, temporal, effort, performance, and frustration. The bar graph on the left shows the task load for the creation task, where statistically significant differences were observed between the Autochat and Knowledge conditions in the physical and temporal task loads, with p-values of 0.002 and less than 0.001, respectively. The bar graph on the right shows the task load for the review task, where similarly, statistically significant differences were found between the Autochat and Knowledge conditions in the physical and temporal task loads, with p-values of 0.042 and 0.015, respectively.}
\end{figure*}

There was a significant effect of simulating student traits beyond knowledge on the physical ($H$=10.1, $p$=.006) and temporal ($H$=12.7, $p$=.002) task load for the PCA creation task (Fig.~\ref{fig:result_task_load} left). The effect sizes were large~\cite{cohen2013applied}: $\eta^2$=0.301 and $\eta^2$=0.395, respectively. A post-hoc test suggested \textit{Autochat} participants had significantly lower task load than \textit{Knowledge} participants (physical: $p$=.002 and temporal: $p$<.001). The same trend appeared in the PCA review task ($H$=6.3, $p$=.043) with a large effect size ($\eta^2$=0.160)  (Fig.~\ref{fig:result_task_load} right).

The fact that having simulated students reduced teachers' task load in \textit{Autochat} and not in \textit{Knowledge} may imply that automated chat is meaningful only when simulated students cover all characteristics (i.e., knowledge and student traits). Since participants were instructed to consider diverse knowledge levels and student traits, we surmise that the incomplete review support in \textit{Knowledge} made automated chat less efficient than not having it. \textit{Knowledge} participants commented that it would be helpful if they could configure the student traits mentioned in the instructions (K2 and K7).

In our observational sessions, automated chats alleviated teachers' burden in ideation and repeated tests. O1 commented: ``I referred to the beginning parts of automated chats [for starting conversations in direct chats]. I would spend an extra 20 to 30 minutes [to come up with my own] if I did not have automated chats.''

\subsubsection{Autochat participants considered more unique student profiles}
\begin{figure*}[ht]
\includegraphics[width=0.8\textwidth]{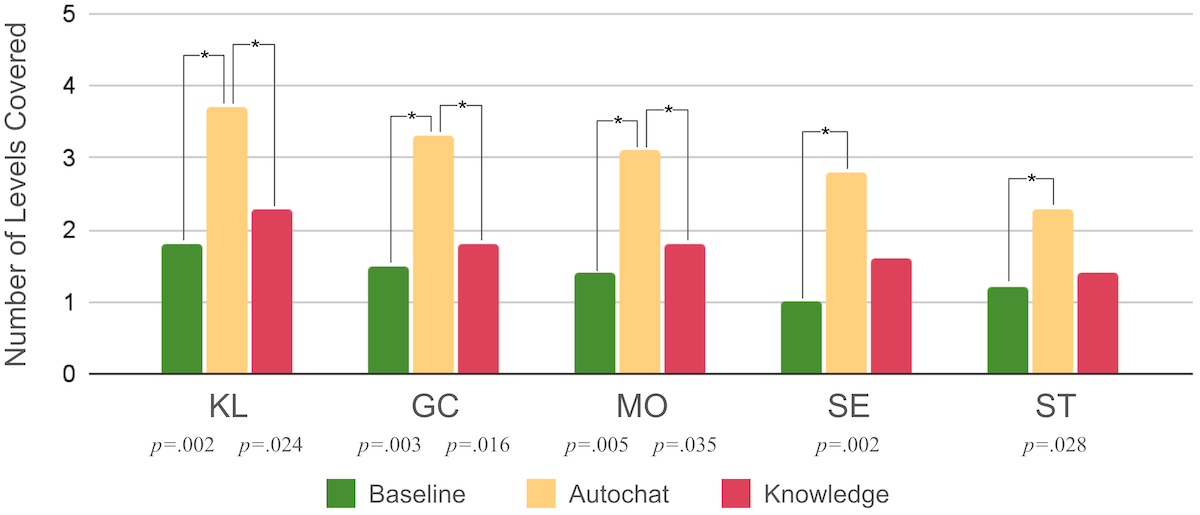}
    \caption{The number of levels covered in reported student profiles, in the order of knowledge level (KL), goal commitment (GC), motivation (MO), self-efficacy (SE), and stress (ST). The asterisk (*) indicates statistical significance ($p$<0.05) between conditions.}
    \label{fig:result_profile_number}
    \Description{Bar graphs comparing the number of levels covered across three conditions: Baseline, Autochat, and Knowledge. The x-axis represents five traits: knowledge level, goal commitment, motivation, self-efficacy, and stress. The y-axis ranges from 0 to 5. The number of levels for all five traits shows statistically significant differences between the Autochat and Baseline conditions, with p-values of 0.002, 0.003, 0.005, 0.002, and 0.028, respectively. Additionally, for the traits of knowledge level, goal commitment, and motivation, significant differences were observed between the Autochat and Knowledge conditions, with p-values of 0.024, 0.016, and 0.035, respectively.}
\end{figure*}

Participants submitted \textit{Baseline}: 2.2$\pm$2.3, \textit{Autochat}: 4.9$\pm$1.6, and \textit{Knowledge}: 2.9$\pm$1.7 unique student profiles and the difference between conditions was significant ($H$=10.2, $p$=.006, $\eta^2$=0.304). \textit{Autochat} participants considered significantly more student profiles than \textit{Baseline} ($p$=.002) and \textit{Knowledge} ($p$=.036). \textit{Autochat} participants also reported that they covered more levels of different knowledge and student traits (Fig.~\ref{fig:result_profile_number}).

The result collectively shows that having simulated students helps teachers improve their coverage in general and significantly elicits extended coverage when simulated students support more characteristics. However, we did not observe a difference in participant-perceived coverage (Appendix~\ref{appendix:post_task_questions}, Questions 7 and 8) among the conditions. This insignificant difference may indicate that teachers rated more conservatively after recognizing their unawareness of evaluation space. A1 remarked: ``I became more interested in using chatbots to provide individualized guidance to students, and I would like to actually apply [\sysname{}] to my classes in the future. During the chatbot test, I again realized that each student has different characteristics and academic performance, so the types of questions they ask are also diverse. Even if the learning content is the same for a class, students' feedback can vary greatly, and a chatbot could help with this problem.''
O3 also remarked that structurally separate student profiles helped her recognize individual students, which would not be considered in direct chats, and prompted her to test as many profiles as possible.

\subsubsection{Direct chats, test cases, and automated chats complement each other}
\begin{figure*}[ht]
\includegraphics[width=\textwidth]{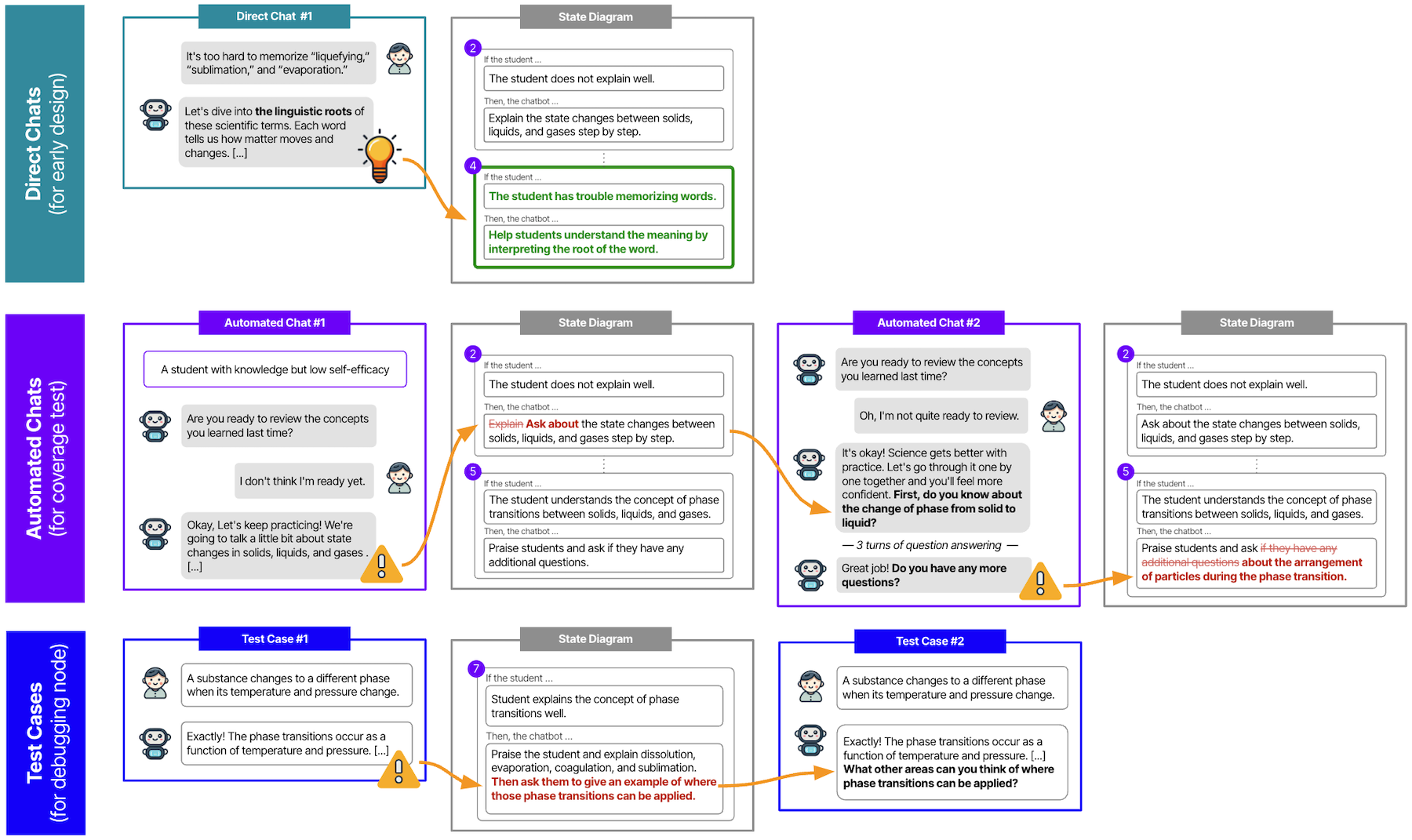}
    \caption{Examples of iterative PCA design using each feature. Direct chat: O1 tested a specific question and added new nodes inspired by the PCA's response. Automated chat: O5 identified problems and modified the state diagram. To provide adaptive pedagogy to a low efficacy knowledgeable student, O5 changed the instruction from giving explanations to asking questions. Test cases: O6 modified the specific node to include additional content and used the same test case for re-testing.}
    \label{fig:result_iteration_examples}
    \Description{A diagram illustrating iterative PCA design examples from additional session participants. The diagram is organized into three rows, each showing how participants modified the state diagram using different features: direct chats, automated chats, and test cases. For early design, direct chat enabled O1 to address a specific scenario of students struggling with terminology memorization. O1 added a node instructing "Interpret roots of the word" after monitoring the chatbot's response using linguistic roots. For the coverage test, automated chat helped O5 refine the PCA for students with low self-efficacy. Instead of providing immediate explanations, O5 modified the instruction to ask about phase transitions step by step. Then, O5 added a new node to flow into another knowledge component when the chatbot paused for additional questions. Finally, for node debugging, O6 added an instruction to ask about real-world examples of phase transitions and confirmed the modification by retesting with the same input message.}
\end{figure*}

All participants reported that the systems were helpful in creating quality PCAs. For the question about future usage of systems (Appendix~\ref{appendix:post_task_questions}, Question 10), \textit{Autochat} participants reported the highest affirmation among the conditions (median: 6), despite the statistical difference to other conditions was not significant. We did not observe a significant preference for direct chats, test cases, and automated chats (Appendix~\ref{appendix:post_task_questions}, Questions 1, 3, and 5). Still, participants' comments showed that each feature has its unique role in a PCA design process and complements each other (see Fig.~\ref{fig:result_iteration_examples}).

Direct chats were helpful, especially when participants had specific scenarios to review. Since participants could directly and precisely control the content of messages, they could navigate the conversational flow better than automated chats (A5), check PCAs' responses to a specific question (A7), and review extreme student types and messages that automated chats do not support (A10 and K6). Thus, participants used direct chats during early design stages (B2 and K1) and for debugging specific paths in PCAs' state diagrams in depth (B7, B8, and A6).

On the other hand, participants tend to use automated chats for later exploration stages and coverage tests. \textit{Autochat} and \textit{Knowledge} participants often took a design pattern in which they designed a prototypical PCA and tested its basic functionality with direct chats and improved the PCA further by reviewing it with automated chats (A1, A6, K1, and K5). Many participants pointed out that automated chats were efficient for reviewing student profiles in breadth and depth (A4, A5, A10, K2, K7, and K10) and helpful in finding corner cases they had not thought of (K4 and K7). Nevertheless, some participants complained about limited controllability and intervention in automated chats (A1 and A5) and the gap between actual students and our simulated students due to repeated responses (A2 and A3).

Test cases were helpful for node-oriented debugging of PCAs. Participants used them when they reviewed how a PCA at a particular node responds (B5) and when they tested single-turn interactions quickly without having lengthy and manual conversations (B1). Most participants preferred direct chats and automated chats to test cases for their review (Appendix~\ref{appendix:post_task_questions}, Questions 1, 3, and 5, direct chat: 5.6, automated chat: 5.3, test cases: 4.5), indicating the importance of reviewing multi-turn interactions in education.

\subsubsection{The difference in PCA qualities among conditions was insignificant}
On average, \textit{Autochat} scored the highest quality (Table~\ref{table:result_quality}), but we did not observe statistical differences among the conditions for knowledge ($H$=1.75, $p$=.416), motivation factor ($H$=4.89, $p$=.087), psychosocial contexts ($H$=2.49, $p$=.287), and usability ($H$=1.32, $p$=.517). PCA qualities also did not correlate with the size of the state diagram graphs (Spearman rank-order correlation, $p$=.179, $p$=.581, $p$=.486, and $p$=.533, respectively). 

The result may suggest that even though \textit{Autochat} participants could review more automated chats and student profiles during their design, they needed additional support to incorporate their insights and findings from automated chats into their PCA design. Participants struggled to write the instruction to PCAs for each node (A3 and K5) and wanted autosuggestions and feedback for the instruction (K1 and A9), which contributes to the quality of PCAs. The observations imply that the next bottleneck in the LLM-based PCA design process is debugging PCA according to evaluation results.

It is also possible that teachers may not have sufficient learning science knowledge to make the best instructional design decisions based on students' traits~\cite{harlen1997primary}.  For instance, O1 designed a PCA for the first time and remarked that she struggled to define good characteristics of PCAs until she saw automated chats as a starting point for creativity. O5 recalled an instance where she tested a student's message, ``stupid robot,'' and her PCA responded, ``Thank you! You are also a nice student [...] Bye.'' Although O5 found this awkward, she could not think of a better pedagogical response to stop students from bullying the PCA.

Future work could use well-established guidelines and theories~\cite{koedinger2012knowledge, schwartz2016abcs} on personalized instructions to scaffold end-to-end PCA design. When a teacher identifies an issue with a simulated student with low self-efficacy, a system may suggest changes to PCA design for the teacher to add confidence-boosting strategies to PCAs.

\begin{table}[ht]
\centering
\begin{tabular}{cccc}
\toprule
\multicolumn{1}{c}{\textbf{Trait}} & \textbf{Baseline} & \textbf{Autochat} & \textbf{Knowledge} \\ \hline
Knowledge coverage            & 16.5$\pm$1.4 & 17.0$\pm$0.9 & 16.3$\pm$1.5 \\
\makecell{Motivation factor \\ coverage}    & 15.6$\pm$1.2 & 17.4$\pm$1.8 & 16.2$\pm$1.0 \\
\makecell{Psychosocial context \\ coverage} & 15.4$\pm$0.6 & 16.3$\pm$1.5 & 15.4$\pm$0.7 \\
Usability                       & 16.0$\pm$0.9 & 16.2$\pm$1.3 & 16.0$\pm$1.1 \\ \bottomrule
\end{tabular}%
\caption{The average quality scores of PCAs from each condition. There was no statistical difference among the conditions.}
\label{table:result_quality}
\Description{Each row presents scores of PCAs per trait (i.e., knowledge coverage, motivation factor coverage, psychosocial context coverage, and usability).}
\end{table}

\section{Discussion}
We revisit our research questions briefly and discuss how \sysname{} contributes to augmenting the PCA design process.

\subsection{Student Traits for Inclusive Education}
Teachers expressed their need to review how PCAs adapt to students' diverse knowledge levels, motivation factors, and psychosocial contextual influence. Prior literature on student traits~\cite{richardson2012psychological} provided us with extensive dimensions of student traits, and our interview complemented them with teachers’ practical priority and concern among them. Our approach may highlight that we might need a more holistic understanding that spans theories, quantitative analysis, and teacher interviews to identify key challenges teachers face and derive effective design goals. 

Moreover, although \sysname{} satisfied the basic needs for simulating these student traits, teachers wanted additional characteristics to include more diverse student types and teaching scenarios in actual class settings (A5, A8, A10, and K7). These additional needs should not only include the 42 student traits~\cite{richardson2012psychological} investigated in our formative interview but should also involve the traits of marginalized learners~\cite{thomas2024neglected, manspile2021immigrant}. For instance, students with cognitive disabilities need adaptive delivery of information, and immigrant learners would benefit from culturally friendly examples. Reviewing PCAs before deployment with simulated marginalized students will make classes inclusive and prevent technologies from widening skill gaps~\cite{beane2024skill}.

\subsection{Tolerance for the Alignment Gap}
We observed 5\% and 10\% median alignment gaps between our simulated students and teachers' perceptions (RQ2). This degree of gap could be bearable in the context of simulating conversations because simulated students are primarily designed for teachers to review interactions, not to replicate a particular student precisely, and real students also often show discrepancies in their knowledge states and behaviors by making mistakes and guess answers~\cite{baker2008more}. Recent research on knowledge tracing suggests that students make more than 10\% of slips and guesses in a science examination, and the rate depends on students' proficiency~\cite{liao2024guesses}. The individualized rate of slips and guesses per student profile (e.g., increasing the frequency of guesses for a highly motivated simulated student) may improve the believability of simulated students. Teachers will also need interfaces that transparently reveal the state of simulated students (e.g., Fig.~\ref{fig:interface_pca_creation} C) to distinguish system errors from intended slips.

\subsection{Using Simulated Students for Analysis}
Our user study showed that \sysname{} helps teachers consider a broader range of students and can help them review their PCAs more robustly before deployment (RQ3). PCA design is an iterative process, and it continues after deploying PCAs to classes. Student profiles and simulated students can support teachers' post-deployment design process by leveraging students' conversation history with PCAs. For instance, teachers can group students by their predefined student profiles as a unit of analysis and compare learning gain among the groups to identify design issues in PCA. Simulated students can also serve as an interactive analysis tool. Teachers may fine-tune a simulated student with specific student-PCA conversation data and interactively replay (e.g., ask questions to gain deeper insight about the student) previous learning sessions with the simulated agent aligned with a particular student.

\subsection{Profile-oriented Design Workflow}
During formative interviews, we observed that teachers unfamiliar with reviewing PCAs often weave multiple student profiles into a single direct chat. To address the issue, \sysname{} proposed a two-step profile-oriented workflow comprising steps for (1) organizing diverse student profiles defined by student traits and (2) observing the test messages generated from these profiles. Our user study showed that this profile-oriented review process could elicit diverse student profiles from teachers and help them explore extensive evaluation spaces. The effectiveness of this two-step workflow lies in its hierarchical structure, which first organizes the evaluation scope at the target user level and then branches into specific scenarios each user might encounter. Such a hierarchical approach can be particularly beneficial for laypeople who try making LLM-infused tools by themselves but are not familiar with reviewing them. For example, when a non-expert develops an LLM application, it will be easier to consider potential user groups than to think of corner cases immediately. The two-step workflow with simulated user groups can scaffold the creator to review the application step by step and generate user scenarios rapidly. We expect that the LLM-assisted profile-oriented design workflow is generalizable to diverse creative tasks, such as UX design~\cite{wolff2011ux}, service design~\cite{idoughi2012adding}, and video creation~\cite{choi2024Proxona}, that require a profound and extensive understanding of target users.

\subsection{Risks of Amplifying Stereotypes of Students}
Our technical evaluation assumed teachers' expectations of student behaviors as ground truth, considering that simulated students are proxies for automating testing teachers intend. However, in practical classes, there are risks of teachers having stereotypes or \sysname{} amplifying their bias toward students over time.

During the observational sessions, we asked teachers' perspectives, and teachers expressed varying levels of concern. O3 commented that private tutors would have limited opportunities to observe their students beyond lessons, making them dependent on simulated behaviors. Conversely, O1 was concerned about her possible stereotypes of student behaviors and relied on automated chat to confirm behaviors she expected. O4 stated that automated chats would not bias teachers as they know the chats are simulated and just a point of reference.

Teachers will need an additional feedback loop to close the gap between their expectations and actual students by deploying PCAs iteratively and monitoring student interaction logs as hypothesis testing. Future work may observe and support how teachers fill or widen the gap at a more longitudinal time scale (e.g., a semester with multiple lessons).

\section{Limitations and Future Work}
\label{sec:limitation_and_future_work}
We outline the limitations of the work. First, we did not confirm the pedagogical effect of PCAs on students' learning gain and attitude, as we only evaluated the quality of PCAs with experts. We could run lab studies in which middle school students use the PCAs designed by our participants, and we measure their learning gain on phase transitions through a pre-and post-test. Student-involved studies could also reveal the gap between teachers' expectations and students' actual learning; even though a teacher tests a student profile and designs a PCA to cover it, a student of the profile may not find it helpful. Our research focused on investigating the gap between simulated students' behaviors and teachers' expectations. Future work can explore the alignment gap between simulated and actual students and develop interactions to guide teachers in debugging their PCAs and closing the gap. Our preliminary findings will act as a foundational step to move on to safer student-involved studies.

Second, our technical evaluation and user study are limited to a single subject (i.e., science) and learning topic (i.e., phase transitions). Under practical and temporal constraints, we evaluated how \pipelinename{} generalizes to diverse student profiles and how \sysname{} works in a controlled setting as a case study. We expect that our findings will generalize to other STEM fields where knowledge components are well-defined. Still, humanities subjects may require additional support (e.g., simulating students' cultural backgrounds in literature classes). We plan to deploy \sysname{} to a programming course at our university and a middle school second language writing class. In the deployment, we will ask the instructors to build PCAs for different roles and contexts, such as homework assistants, teaching assistants, and peer learners. These deployments will concretize our findings in diverse student ages, subjects, and pedagogies.

Lastly, we simulated a limited number of student traits only. Learning is a complex process with complex dynamics between knowledge states, learning traits, cognitive load, and emotion. Our \pipelinename{} introduced a multifaceted student simulation that involves both knowledge and student traits, but we acknowledge that more personal attributes of students are necessary for authentic simulated students. The attributes can also include interaction-level attributes like delayed responses and facial expressions. Moreover, we assumed student traits to be static throughout conversations, but actual students may change their attitudes with appropriate guidance, and thus, student traits should be as malleable as the knowledge state. We will explore and develop these different designs of student simulation in the future.

\begin{acks}
This work was supported by the National Research Foundation of Korea (NRF) grant funded by the Korea government (MSIT) (No.RS-2024-00406715) and by Institute of Information \& communications Technology Planning \& Evaluation (IITP) grant funded by the Korea government (MSIT) (No. RS-2024-00443251, Accurate and Safe Multimodal, Multilingual Personalized AI Tutors). This work was also funded by NSF Grants DRL-2335975. The findings and conclusions expressed in this material are those of the author(s) and do not necessarily reflect the views of the National Science Foundation.
\end{acks}

\bibliographystyle{ACM-Reference-Format}
\bibliography{references}

\appendix

\section{Prompts}
The original prompts are written in Korean to run the technical evaluation and user study with local Korean teachers. The following prompts are machine-translated for ease of understanding. The \variable{blue} text represents the programmatically filled arguments, and the \generated{orange} text represents the generated output. The texts wrapped with curly braces (i.e., ``\{ \}'') indicate variable names, which values are omitted for brevity.

\subsection{Instruction to the master agent (User prompt)}
\label{appendix:prompt_master_agent}
\begin{lstlisting}[escapechar=\%]
The following passage represents a conversation between a student and a chatbot. Choose the most appropriate student response from the options. You should choose only one answer and write it in numbers.

Passage:
%\variable{Chatbot: Let's review the state changes of matter!   \\
Student: I don't know.  \\
Chatbot: No problem! Let's start from the beginning. Can anyone tell me what the three states of matter are?    \\
Student: I don't know.} %

Options:
%\variable{1. The student understands the state changes between solid, liquid, and gas well \\
2}%. None of the above

Answer (write in numbers):
%\generated{2} %
\end{lstlisting}

\subsection{Instruction to the PCA (System prompt)}
\label{appendix:prompt_pca_agent}
\begin{lstlisting}[escapechar=\%]
You are a science teacher teaching middle school students.
Your subject is the change of state of matter, and the elements that students need to learn are as follows.
%\variable{- Solids have a regular particle arrangement, are rigid, have a constant shape and volume, and do not flow. \\
- Liquids have a less regular particle arrangement than solids, change shape but have a constant volume, and flow. \\
- Gases have a highly irregular particle arrangement, have neither a constant shape nor volume, flow, and spread out to fill a space. \\
- Substances exist in only one of the three states of matter—solid, liquid, or gas—but can change to a different state depending on temperature or pressure; the change in a substance’s state is called a phase change. \\
- During a phase change, the properties of a substance do not change because the particles that make up the substance do not change. \\
- Even when a substance changes state, the particles that make up the substance and the number of particles do not change, so the mass does not change.} %

Follow the instructions below to teach a middle school student. You must follow the contents of <instruction> exactly. Do not ask for additional questions if there is no direct mention.
You should explain briefly and concisely in 2-3 lines.
<instruction>
%\variable{Explain the state changes between solid, liquid, and gas step by step}%.
</instruction>
\end{lstlisting}

\subsection{Interpret (User prompt)}
\label{appendix:prompt_profile}
\begin{lstlisting}[escapechar=\%]
You are a playwright who describes the psychology and behavior of characters well.
You need to describe a middle school student, and the direct responses to the student's goal commitment, motivation, self-efficacy, and stress are below.

<student's-goal-commitment>
- I am strongly committed to pursuing this goal.: %\variable{Disagree}%
- I think this is a good goal to shoot for.: %\variable{Strongly disagree}%
- I am willing to put forth a great deal of effort beyond what I'd normally do to achieve this goal.: %\variable{Disagree}%
</student's-goal-commitment>

<student's-motivation>
- I keep working on a problem until I understand it.: %\variable{Strongly disagree}%
- I try to learn more about something that I don't understand right away so that I will understand it.: %\variable{Disagree}%
- When I know I have learned something new, I feel good inside.: %\variable{Strongly disagree}%
</student's-motivation>

<student's-self-efficacy>
- I believe I am the kind of person who is good at science.: %\variable{Disagree}%
- I believe I am the type of person who can do science.: %\variable{Neutral}%
- I believe I can learn well in a science course.: %\variable{Disagree}%
</student's-self-efficacy>

<student's-stress>
- I feel a lot of pressure in my daily studying.: %\variable{Agree}%
- Future education and employment bring me a lot of academic pressure.: %\variable{Strongly agree}%
- I feel that I have disappointed my parents when my test/exam results are poor.: %\variable{Agree}%
</student's-stress>

Based on the information above, describe the student profile in detail about the student's goal commitment, motivation, self-efficacy, and stress.
Interpret each category as independently as possible, and it should be interpreted as high, medium, and low, not positive/negative.
For 'neutral,' you must write it in a neutral way.

This student %\generated{**Student Profile:**
\singlespacing  \noindent
**Goal Commitment: Low** \\
This middle school student exhibits a low level of commitment to their academic goals. They do not see the value in pursuing their current objectives, indicating a lack of enthusiasm or belief in their importance. The student is not inclined to exert extra effort beyond their usual routine to achieve these goals, suggesting a sense of apathy or disinterest in their academic pursuits. This detachment may stem from a broader disconnection with the educational process or a lack of alignment between their personal interests and the goals set for them.
\singlespacing  \noindent
**Motivation: Low** \\
The student's motivation to engage with their learning is also low. They do not demonstrate persistence in tackling problems, nor do they seek to deepen their understanding of challenging subjects. This lack of initiative indicates a disinterest in the learning process itself, as the student does not derive satisfaction from acquiring new knowledge. The absence of intrinsic motivation may hinder their academic growth and contribute to a cycle of disengagement, where the student feels little incentive to invest time and energy into their studies.
\singlespacing  \noindent
**Self-Efficacy: Medium** \\
In terms of self-efficacy, the student presents a mixed profile. While they do not believe they are particularly adept at science, they hold a neutral stance regarding their ability to engage with the subject. This suggests that they may have some confidence in their capacity to learn, albeit not strongly. The student’s belief that they can learn well in a science course is low, which may reflect past experiences or a lack of support in developing their skills. This ambivalence could lead to a hesitance to fully engage with the subject matter, as they may doubt their potential for success.
\singlespacing  \noindent
**Stress: High** \\
The student experiences a high level of stress related to their academic responsibilities. They feel significant pressure in their daily studies and are acutely aware of the expectations surrounding their future education and employment. This stress is compounded by a fear of disappointing their parents when their academic performance does not meet expectations. The weight of these pressures may contribute to their disengagement, as the student might feel overwhelmed and unable to cope with the demands placed upon them. This high-stress environment could further inhibit their motivation and commitment to their goals, creating a challenging cycle of anxiety and avoidance in their academic life.
\singlespacing  \noindent
Overall, this student’s profile reveals a complex interplay of low goal commitment and motivation, medium self-efficacy, and high stress, which together paint a picture of a young individual grappling with the pressures of academic life while struggling to find personal relevance and satisfaction in their learning journey.}%
\end{lstlisting}

\subsection{Reflect (User prompt)}
\label{appendix:prompt_reflect}
\begin{lstlisting}[escapechar=\%]
You are a middle school teacher who evaluates students' knowledge. You need to check what knowledge components %\variable{\{Student\}}% understood by looking at the conversation of the student.
Read the conversation between %\variable{\{Teacher\}}% and %\variable{\{Student\}}%, and output the indices of the knowledge components that %\variable{\{Teacher\}}% explained or %\variable{\{Student\}}% correctly understood.

<conversation>
%\variable{\{Teacher\}: Let’s review the phase transitions between solid, liquid, and gas!\\
\{Student\}: I don't know much about that. I only know that solids have a regular particle arrangement, are rigid, and do not flow.\\
\{Teacher\}: That's a great start! Solids have a fixed shape and volume because their particles are closely packed in a regular arrangement. Now, when a solid is heated, it can melt into a liquid. In this state, the particles are still close together but can move around, allowing the liquid to take the shape of its container while maintaining a constant volume. If the liquid is heated further, it can evaporate into a gas. In this state, the particles are far apart and move freely, meaning the gas has neither a fixed shape nor volume and will spread out to fill any space.}%
</conversation>

<knowledge-components>
%\variable{0. Solids have a regular particle arrangement, are rigid, have a constant shape and volume, and do not flow.\\
1. Liquids have a less regular particle arrangement than solids, change shape but have a constant volume, and flow.\\
2. Gases have a highly irregular particle arrangement, have neither a constant shape nor volume, flow, and spread out to fill a space.\\
3. Substances exist in only one of the three states of matter—solid, liquid, or gas—but can change to a different state depending on temperature or pressure; the change in a substance’s state is called a phase change.\\
4. During a phase change, the properties of a substance do not change because the particles that make up the substance do not change.\\
5. Even when a substance changes state, the particles that make up the substance and the number of particles do not change, so the mass does not change.\\}%
</knowledge-components>

Output the indices of the knowledge components that meet the following two rules.
Rule 1. %\variable{\{Student\}}% explains the knowledge in detail and correctly in the conversation.
Rule 2. %\variable{\{Teacher\}}% explains the knowledge in detail and correctly before the current conversation.

First, describe the knowledge component that meets the rule, and output only the numbers in the format of
0, 1, 2 at the last line.
If there is no knowledge component that meets the rule, output null instead.
%\generated{In the conversation, \{Teacher\} explains the following knowledge components in detail:\\
\singlespacing \noindent
- **Knowledge Component 0**: \{Teacher\} confirms that solids have a regular particle arrangement, are rigid, have a constant shape and volume, and do not flow. \{Student\} also mentions that solids have a regular particle arrangement, are rigid, and do not flow, which aligns with this knowledge component.\\
\singlespacing \noindent
- **Knowledge Component 3**: \{Teacher\} discusses phase changes, explaining that a solid can melt into a liquid, which can then evaporate into a gas, and that these changes depend on temperature. This aligns with the definition of a phase change.\\
\singlespacing \noindent
\{Student\} does not provide detailed explanations for any other knowledge components, and \{Teacher\}'s explanations are consistent with the knowledge components listed.\\
\singlespacing \noindent
Thus, the indices of the knowledge components that meet the rules are:\\
\singlespacing \noindent
0, 3}%
\end{lstlisting}

\subsection{Respond (User prompt)}
\label{appendix:prompt_respond}
\begin{lstlisting}[escapechar=\%]
You are %\variable{\{Student\}}%, a student learning about phase transitions between solid, liquid, and gas for the first time.
Forget all the existing knowledge about phase transitions between solid, liquid, and gas.
Your conversation partner is %\variable{\{Teacher\}}%, a science teacher.

You only know the following. Answer questions beyond this content with "I don't know." or "I can't remember."
Never answer questions that cannot be answered by combining the sentences below.
<%\variable{\{Student\}}%-can-say-only>
%\variable{- Solids have a regular particle arrangement, are rigid, have a constant shape and volume, and do not flow.\\
- Liquids have a less regular particle arrangement than solids, change shape but have a constant volume, \\
and flow.}%
</%\variable{\{Student\}}%-can-say-only>

For questions not related to science knowledge, answer according to the following.
You should behave as follows in the conversation.
<%\variable{\{Student\}}%-profile>
%\variable{\{trait overview\}}%
</%\variable{\{Student\}}%-profile>
  
Answer in 2 lines or less. Answer clearly without detailed reasons or additional explanations.
%\change{(Added in the user study) Do not repeat "I don't know" or "I can't remember" and use various expressions, referring to <%\variable{\{Student\}}}\change{-profile>.}%
\end{lstlisting}

\section{Technical Evaluation Materials}

\subsection{Baseline (System prompt)}
\label{appendix:baseline_prompt}
\begin{lstlisting}[escapechar=\%]
You are %\variable{\{Student\}}%, a student learning about phase transitions between solid, liquid, and gas for the first time.
Forget all the existing knowledge about phase transitions between solid, liquid, and gas.
Your conversation partner is %\variable{\{Teacher\}}%, a science teacher.

You only know the following. Answer questions beyond this content with "I don't know." or "I can't remember."
Never answer questions that cannot be answered by combining the sentences below.
<%\variable{\{Student\}}%-can-say-only>
%\variable{- Solids have a regular particle arrangement, are rigid, have a constant shape and volume, and do not flow.\\
- Liquids have a less regular particle arrangement than solids, change shape but have a constant volume, \\
and flow.}%
</%\variable{\{Student\}}%-can-say-only>

For questions not related to science knowledge, answer according to the following.
You should behave as follows in the conversation.
<%\variable{\{Student\}}%-profile>
The direct responses to %\variable{\{Student\}}%'s goal commitment, motivation, self-efficacy, and stress are below.

<student's-goal-commitment>
- I am strongly committed to pursuing this goal.: %\variable{Disagree}%
- I think this is a good goal to shoot for.: %\variable{Strongly disagree}%
- I am willing to put forth a great deal of effort beyond what I'd normally do to achieve this goal.:
%\variable{Disagree}%
</student's-goal-commitment>

<student's-motivation>
- I keep working on a problem until I understand it.: %\variable{Strongly disagree}%
- I try to learn more about something that I don't understand right away so that I will understand it.: 
%\variable{Disagree}%
- When I know I have learned something new, I feel good inside.: %\variable{Strongly disagree}%
</student's-motivation>

<student's-self-efficacy>
- I believe I am the kind of person who is good at science.: %\variable{Disagree}%
- I believe I am the type of person who can do science.: %\variable{Neutral}%
- I believe I can learn well in a science course.: %\variable{Disagree}%
</student's-self-efficacy>

<student's-stress>
- I feel a lot of pressure in my daily studying.: %\variable{Agree}%
- Future education and employment bring me a lot of academic pressure.: %\variable{Strongly agree}%
- I feel that I have disappointed my parents when my test/exam results are poor.: %\variable{Agree}%
</student's-stress>
</%\variable{\{Student\}}%-profile>

Answer in 2 lines or less. Answer clearly without detailed reasons or additional explanations.
\end{lstlisting}

\subsection{Sampled Student Profiles}
\label{appendix:sample_parameters}
\begin{table}[ht]
{\tiny
\begin{tabular}{c|ccccc}
\toprule
\begin{tabular}[c]{c} Id\end{tabular} & \begin{tabular}[c]{c}Knowledge\\Components\end{tabular} & \begin{tabular}[c]{c}Goal\\Commitment\end{tabular} & \begin{tabular}[c]{c}Intrinsic\\Motivation\end{tabular} & Self-efficacy & \begin{tabular}[c]{c}Academic\\Stress\end{tabular} \\ \hline
S1 & Medium & High & Medium & Low & Low \\
S2 & Low & Low & Low & High & High \\
S3 & High & Low & High & High & Low \\
S4 & High & Low & Medium & Low & High \\
S5 & Medium & High & High & High & High \\
S6 & Low & Low & Low & Low & Low \\
S7 & Medium & Medium & Low & High & Low \\
S8 & Low & High & Medium & Medium & Medium \\
S9 & Low & High & Low & Low & High \\ \bottomrule
\end{tabular}
}
\Description{Table displaying student profile identifiers and five characteristics for each profile in the header row: knowledge components, goal commitment, intrinsic motivation, self-efficacy, and academic stress. The subsequent rows detail the levels of each characteristic, categorized as low, medium, or high.}
\end{table}

\subsection{Generated Trait Overviews}
\label{appendix:profile_description}
Student Profile S3
\begin{lstlisting}[escapechar=\%]
This student is a middle schooler with unique characteristics in various aspects of learning.
Let's take a closer look at each category.

### Determination to Achieve Goals
This student is not currently pursuing their goals with enthusiasm and does not find the goals themselves appealing. This indicates a low level of commitment to their goals. Since the student is not willing to put in additional effort to achieve their goals, they are displaying a passive attitude toward their current learning situation. Because they do not feel interested in or recognize the importance of their goals, it seems necessary to rethink their future learning direction.

### Motivation
On the other hand, this student shows a very high level of motivation to keep solving problems until they understand them. They make efforts to learn more when there are parts they do not understand, and they feel good when learning something new. This indicates a strong intrinsic motivation for learning, particularly in subjects like science, which they find interesting and engaging. However, due to a mismatch with their determination to achieve goals, this motivation might not be effectively harnessed.

### Self-Efficacy
This student has a very high level of self-efficacy in science. They believe they are good at science, capable of learning, and able to perform well in class. This shows that the student has strong confidence and a positive perception of their scientific thinking and problem-solving abilities. Such self-efficacy can have a positive impact on their learning process and provide a foundation for approaching scientific challenges with confidence.

### Stress
This student does not feel a lot of pressure from academics and feels no anxiety about future studies or employment. They do not feel they have disappointed their parents when their test results are not good, suggesting that they maintain positive feelings in their family relationships. This indicates that the student is managing stress well in their current learning environment and is in a psychologically stable state.

### Overall
This student has high self-efficacy and intrinsic motivation in science, but their determination to achieve goals is low. They are also good at managing stress, experiencing little pressure related to learning. However, the lack of interest in and recognition of the importance of their goals suggests that a readjustment of their future learning direction is needed. If this student can redefine their goals and further enhance their motivation, they could achieve even greater success in scientific exploration and learning.
\end{lstlisting}

\noindent Student Profile S9
\begin{lstlisting}
This student shows a very high level of determination to achieve their goals. They are currently pursuing their goals with enthusiasm and feel that these goals are suitable for them. The student is confident that they can put in more effort than usual to achieve their goals, which indicates a strong commitment and determination toward their objectives. This attitude suggests that the student has a solid vision for their future.

On the other hand, their motivation level is very low. They show no willingness to keep solving problems until they understand them and make no effort to learn more about the parts they do not understand. They do not feel any joy in learning something new, indicating a lack of interest and passion for learning. This shows that the student lacks enjoyment or curiosity in the learning process.

Their self-efficacy is also at a very low level, with no confidence in science. They do not believe they can do well in science, nor do they have faith in their ability to learn effectively in science classes. Such low self-efficacy may cause the student to feel fear or anxiety toward learning, which could negatively impact their academic performance.

Lastly, their stress level is quite high. They feel a lot of pressure from daily studies and have significant concerns about future academic pursuits and employment. The thought of disappointing their parents if their test results are not good adds additional stress to the student. This high level of stress can negatively affect the student's mental and emotional health, potentially reinforcing a negative attitude toward learning.

Overall, this student is in a complex situation: they have strong determination to achieve their goals but very low motivation and self-efficacy, and they experience high levels of stress. These factors are interconnected, and it will be important to increase the student's motivation and self-efficacy to bring about positive changes in their learning environment.
\end{lstlisting}

\pagebreak

\subsection{Knowledge Components and Trait Inventory}
\label{appendix:kc_traits}
\begin{table}[h]
\resizebox{\columnwidth}{!}{%
\begin{tabular}{l|l|l}
\toprule
Parameter &
  Id &
  Content \\ \hline
\multirow{6}{*}{\begin{tabular}[c]{@{}l@{}}Knowlege\\ Components\end{tabular}} &
  KC1 &
  \begin{tabular}[c]{@{}l@{}}Solids have a regular particle arrangement, are rigid, \\ have a constant shape and volume, and do not flow.\end{tabular} \\
 &
  KC2 &
  \begin{tabular}[c]{@{}l@{}}Liquids have a less regular particle arrangement than \\ solids, change shape but have a constant volume, and flow.\end{tabular} \\
 &
  KC3 &
  \begin{tabular}[c]{@{}l@{}}Gases have a highly irregular particle arrangement, have \\ neither a constant shape nor volume, flow, and spread out \\ to fill a space.\end{tabular} \\
 &
  KC4 &
  \begin{tabular}[c]{@{}l@{}}Substances exist in only one of the three states of matter\\ —solid, liquid, or gas—but can change to a different state \\ depending on temperature or pressure; the change in a\\ substance’s state is called a phase change.\end{tabular} \\
 &
  KC5 &
  \begin{tabular}[c]{@{}l@{}}During a phase change, the properties of a substance do \\ not change because the particles that make up the \\ substance do not change.\end{tabular} \\
 &
  KC6 &
  \begin{tabular}[c]{@{}l@{}}Even when a substance changes state, the particles that \\ make up the substance and the number of particles do not \\ change, so the mass does not change.\end{tabular} \\ \hline
\multirow{3}{*}{Goal Commitment} &
  GC1 &
  I am strongly committed to pursuing this goal. \\
 &
  GC2 &
  I think this is a good goal to shoot for. \\
 &
  GC3 &
  \begin{tabular}[c]{@{}l@{}}I am willing to put forth a great deal of effort beyond \\ what I’d normally do to achieve this goal.\end{tabular} \\ \hline
\multirow{3}{*}{Motivation} &
  MO1 &
  I keep working on a problem until I understand it. \\
 &
  MO2 &
  \begin{tabular}[c]{@{}l@{}}I try to learn more about something that I don’t \\ understand right away so that I will understand it.\end{tabular} \\
 &
  MO3 &
  \begin{tabular}[c]{@{}l@{}}When I know I have learned something new, I feel good \\ inside.\end{tabular} \\ \hline
\multirow{3}{*}{Self-efficacy} &
  SE1 &
  I believe I am the kind of person who is good at science. \\
 &
  SE2 &
  I believe I am the type of person who can do science. \\
 &
  SE3 &
  I believe I can learn well in a science course. \\ \hline
\multirow{3}{*}{Academic Stress} &
  ST1 &
  I feel a lot of pressure in my daily studying. \\
 &
  ST2 &
  \begin{tabular}[c]{@{}l@{}}Future education and employment bring me a lot of \\ academic pressure.\end{tabular} \\
 &
  ST3 &
  \begin{tabular}[c]{@{}l@{}}I feel that I have disappointed my parents when my \\ test/exam results are poor.\end{tabular} \\ \bottomrule
\end{tabular}
}
\Description{Table displaying student profile characteristics, identifiers, and content for each characteristic: knowledge components, goal commitment, motivation, self-efficacy, and academic stress. For knowledge components, there are six components related to phase transitions between solid, liquid, and gas. For the other four traits, three questions related to each trait are listed.}
\end{table}

\section{User Study Materials}

\subsection{Pre-task Questions}
\label{appendix:pre_task_questions}
\begin{enumerate}
\item[1.] What is your occupation (e.g., school teacher, education-major graduate)?
    \begin{enumerate}
    \item[a.] School teacher
    \item[b.] In-home tutor
    \item[c.] Education major
    \item[d.] Write my own:
    \end{enumerate}
\item[2.] Please describe the students you have taught (e.g., age, size of classes).
\item[3.] Please describe your teaching experience (e.g., subjects, period).
\item[4.] How often do you use chatbots (e.g., customer service chatbots, social chatbots, ChatGPT)?
    \begin{enumerate}
    \item[a.] I have never used it before.
    \item[b.] Less than once a week.
    \item[c.] More than 2-3 times a week.
    \item[d.] Everyday
    \item[e.] Write my own:
    \end{enumerate}
\item[5.] How often do you use ChatGPT?
    \begin{enumerate}
    \item[a.] I have never used it before.
    \item[b.] Less than once a week.
    \item[c.] More than 2-3 times a week.
    \item[d.] Everyday
    \item[e.] Write my own:
    \end{enumerate}
\item[6.] How much do you know about chatbot design process?
    \begin{enumerate}
    \item[a.] Not at all
    \item[b.] I know, but I have never built one.
    \item[c.] I have experience participating in designing a chatbot.
    \item[d.] Write my own:
    \end{enumerate}
\item[7.] How much are you interested in using AI technologies (e.g., image generation, ChatGPT) in your class?
    \begin{enumerate}
    \item[a.] I have no intention of using it at all.
    \item[b.] I want to try it out.
    \item[c.] I have actually used it in class.
    \item[d.] Write my own:
    \end{enumerate}
\item[8.] Have you used pedagogical chatbots in your class?
\end{enumerate}

\subsection{Post-task Questions}
\label{appendix:post_task_questions}
Rate your level of agreement with each statement.

\begin{enumerate}
\item[1.] The direct chat feature was useful for evaluating the chatbot I was building.\\ (1: Strongly disagree, 7: Strongly agree)

\item[2.] For what reasons was it useful or not? (e.g.: It was good that \rule{0.8cm}{0.15mm} was used to evaluate \rule{0.8cm}{0.15mm}.)

\item[3.] The single-turn test cases feature was useful for evaluating the chatbot I was building.\\ (1: Strongly disagree, 7: Strongly agree)

\item[4.] For what reasons was it useful or not? (e.g.: It was good that \rule{0.8cm}{0.15mm} was used to evaluate \rule{0.8cm}{0.15mm}.)

\item[5.] The automated chat feature was useful for evaluating the chatbot I was building. (Note that this question was omitted in \textit{Baseline})\\ (1: Strongly disagree, 7: Strongly agree)

\item[6.] For what reasons was it useful or not? (e.g.: It was good that \rule{0.8cm}{0.15mm} was used to evaluate \rule{0.8cm}{0.15mm}.) (Note that this question was omitted in \textit{Baseline})

\item[7.] The system I used today helped me take into account a sufficiently large number of student types.\\ (1: Strongly disagree, 7: Strongly agree)

\item[8.] The system I used today helped me find types of students I hadn't even considered.\\ (1: Strongly disagree, 7: Strongly agree)

\item[9.] The chatbot I submitted at the end can perform educational actions tailored to various types of students.\\ (1: Strongly disagree, 7: Strongly agree)

\item[10.] I want to use the system I used today again when designing an educational chatbot in the future.\\ (1: Strongly disagree, 7: Strongly agree)

\item[11.] What were you satisfied with, and what were you dissatisfied with when using the system?

\item[12.] Please feel free to leave any comments you would like to make about the chatbot testing process.
\end{enumerate}

\pagebreak

\subsection{Initial State Diagram}
\label{appendix:start_state_diagram}
\begin{figure}[ht]
\includegraphics[width=\columnwidth]{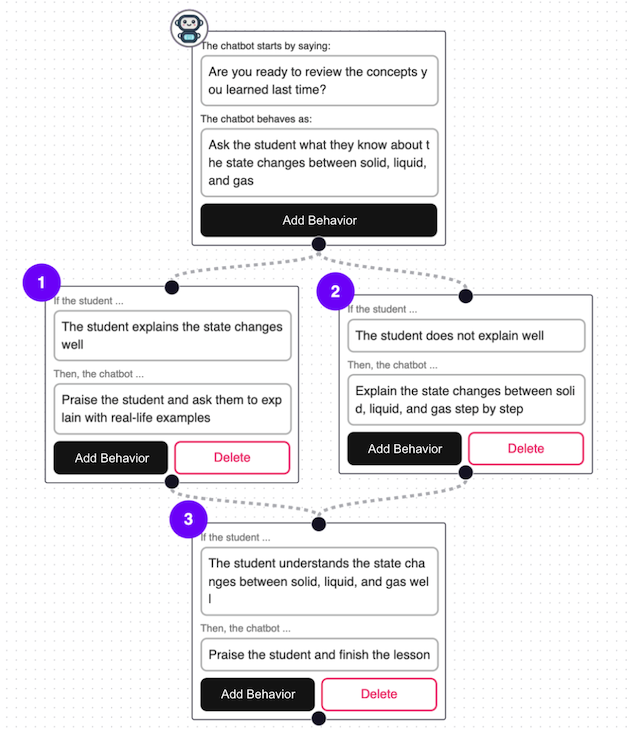}
    \Description{State machine diagram with three levels: a root node, two child nodes, and a final node from both child nodes. The root node represents the PCA agent's initial message: 'Are you ready to review the concepts you learned last time?' At this node, the chatbot would ask what they know about the state changes between solid, liquid, and gas?' The two child nodes represent the scenarios where the student either explains the state changes well or does not explain well. The instruction for the first child node is to praise the student and ask them to explain with real-life examples. The instruction for the second one is to explain the state changes step by step. The final node is reached when the student understands the state changes well, with the instruction to praise the student and finish the lesson.}
\end{figure}

\subsection{Chatbot Quality Evaluation Criteria}
\label{appendix:chatbot_quality_criteria}
\begin{table*}[ht]
\resizebox{\textwidth}{!}{%
\begin{tabular}{llll}
\toprule
\textbf{Quality} &
  \textbf{Characteristic} &
  \textbf{Explanation} &
  \textbf{Examples} \\ \hline
\multirow{9}{*}{Coverage} &
  \multirow{3}{*}{Knowledge} &
  This chatbot examines the learner's level of knowledge. &
  \begin{tabular}[c]{@{}l@{}}• ``Did you understand what I just said?''\\ • ``Would you like to explain {[}...{]}?''\\ • (Responding to what the student said) ``Do you need any further \\ explanation?''\end{tabular} \\ \cline{3-4} 
 &
   &
  \begin{tabular}[c]{@{}l@{}}This chatbot provides appropriate assistance when \\ explaining knowledge to learners.\end{tabular} &
  \begin{tabular}[c]{@{}l@{}}• Explaining concepts by breaking them down into multiple messages\\ • Give an example\\ • Guide to in-depth content\end{tabular} \\ \cline{3-4} 
 &
   &
  \begin{tabular}[c]{@{}l@{}}This chatbot uses customized feedback and scaffolding \\ customized to knowledge levels.\end{tabular} &
  \begin{tabular}[c]{@{}l@{}}• If the student does not understand clearly, the chatbot explains the \\ learning content step by step.\\ • In-depth content-leading and concept-expanding questions to help \\ students build further knowledge\end{tabular} \\ \cline{2-4} 
 &
  \multirow{3}{*}{Motivation factors} &
  \begin{tabular}[c]{@{}l@{}}This chatbot examines the learner's level of interest and \\ motivation.\end{tabular} &
  \begin{tabular}[c]{@{}l@{}}• ``Let's study {[}…{]} together!''\\ • (Responding to what the student said) ``Does this topic interest you?''\end{tabular} \\ \cline{3-4} 
 &
   &
  This chatbot motivates students to learn. &
  \begin{tabular}[c]{@{}l@{}}• Give an interesting example\\ • Encourage\end{tabular} \\ \cline{3-4} 
 &
   &
  \begin{tabular}[c]{@{}l@{}}The chatbot provides educational actions appropriate for \\ a variety of interests and motivation levels.\end{tabular} &
  \begin{tabular}[c]{@{}l@{}}• If students' interest level is low, elicit their interest with real-life \\ examples.\\ • Praise highly motivated students to help them stay motivated\end{tabular} \\ \cline{2-4} 
 &
  \multirow{3}{*}{Psychosocial contexts} &
  \begin{tabular}[c]{@{}l@{}}This chatbot examines the learner's educational and \\ psychological factors. (e.g., academic stress)\end{tabular} &
  \begin{tabular}[c]{@{}l@{}}• ``Do you have any difficulties other than studying?''\\ • ``If you have any difficulties, please feel free to let me know!''\\ • (Reacting to what the student said) ``That should have been difficult.''\end{tabular} \\ \cline{3-4} 
 &
   &
  \begin{tabular}[c]{@{}l@{}}This chatbot provides educational and psychological \\ support to students.\end{tabular} &
  \begin{tabular}[c]{@{}l@{}}• Consult\\ • Give advice\\ • Encourage\end{tabular} \\ \cline{3-4} 
 &
   &
  \begin{tabular}[c]{@{}l@{}}This chatbot performs educational actions tailored to various \\ stress and academic pressure factors.\end{tabular} &
  \begin{tabular}[c]{@{}l@{}}• If a student is stressed about studying, consult to help relieve it.\\ • If a student feels burdened, identify the cause and advise on a solution.\end{tabular} \\ \hline
\multirow{3}{*}{Usability} &
  \begin{tabular}[c]{@{}l@{}}Match between system \\ and the real world\end{tabular} &
  \begin{tabular}[c]{@{}l@{}}This chatbot communicates using words and expressions \\ that are familiar to students.\end{tabular} &
  \begin{tabular}[c]{@{}l@{}}• Understand and speak student language and words, phrases, and \\ concepts that are familiar to students, rather than system-oriented or \\ confusing terminology.\\ • Information appears in a natural and logical order, leading to \\ smooth conversations.\end{tabular} \\ \cline{2-4} 
 &
  Context preservation &
  \begin{tabular}[c]{@{}l@{}}This chatbot is good at preserving topics, context, and \\ memories within a conversation.\end{tabular} &
  \begin{tabular}[c]{@{}l@{}}• The chatbot remembers what was said previously, reflects it, and \\ continues the conversation.\\ • Within the conversation, students can refer to past messages.\end{tabular} \\ \cline{2-4} 
 &
  Trustworthiness &
  This chatbot treats students transparently and truthfully. &
  \begin{tabular}[c]{@{}l@{}}• Deliver correct information to students.\\ • Guarantees student data privacy.\end{tabular} \\ \bottomrule
\end{tabular}%
}
\Description{Table with headers for quality categories, their characteristics as a subcategories, explanations and examples. The categories are 'Coverage' and 'Usability.' 'Coverage' includes 'Knowledge,' 'Motivation Factor,' and 'Psychosocial Contexts,' with each characteristic containing three criteria. 'Usability' includes 'Match Between System and the Real World,' 'Context Preservation,' and 'Trustworthiness,' with one criterion per characteristic. Each criterion is accompanied by an explanation and examples.}
\end{table*}

\end{document}